\documentclass[fleqn,usenatbib]{mnras}
\usepackage{newtxtext,newtxmath}
\usepackage[T1]{fontenc}

\DeclareRobustCommand{\VAN}[3]{#2}
\let\VANthebibliography\thebibliography
\def\thebibliography{\DeclareRobustCommand{\VAN}[3]{##3}\VANthebibliography}

\usepackage{graphicx}	
\usepackage{amsmath}	
\usepackage[dvipsnames]{xcolor}
\usepackage{enumitem}
\usepackage{url}
\usepackage{braket}
\usepackage{threeparttable}
\usepackage{hyperref}
\usepackage{multirow, tabularx}

\def\revstyle{}
\def\term[#1]{{\bf \ttfamily #1}}
\def\B[#1]{{\bf #1}}

\setlist[itemize,1]{label=$\bullet$}
\setlist[itemize,2]{label=$\bullet$}
\setlist[itemize,3]{label=$\bullet$}
\setlist[itemize,4]{label=$\bullet$}
\setlist[itemize]{leftmargin=*}
\def\bit{\begin{itemize}[topsep=0em,parsep=0em,itemsep=0em,partopsep=0em,leftmargin=*]}
\def\bitt{\begin{itemize}[topsep=0em,parsep=0em,itemsep=0em,partopsep=0em,leftmargin=3.0em]}
\def\eit{\end{itemize}}
\def\benum{\begin{enumerate}[leftmargin=1em,itemsep=0pt,parsep=0pt,topsep=0pt]}
\def\eenum{\end{enumerate}}

\def\beq{\begin{equation}}
\def\eeq{\end{equation}}
\def\bey{\begin{eqnarray}}
\def\eey{\end{eqnarray}}

\def\bfrm[#1]{\mathrm{{\bf#1}}}

\def\gs{\mathrel{\raise1.16pt\hbox{$>$}\kern-7.0pt
        \lower3.06pt\hbox{{$\scriptstyle \sim$}}}}
\def\ls{\mathrel{\raise1.16pt\hbox{$<$}\kern-7.0pt
        \lower3.06pt\hbox{{$\scriptstyle \sim$}}}}

\def\gtsima{\, {\buildrel > \over \sim} \,}
\def\ltsima{\, {\buildrel < \over \sim} \,}
\def\prosima{\, {\buildrel \propto \over \sim} \,}
\def\gsim{\lower.5ex\hbox{\gtsima}}
\def\lsim{\lower.5ex\hbox{\ltsima}}
\def\simgt{\lower.5ex\hbox{\gtsima}}
\def\simlt{\lower.5ex\hbox{\ltsima}}
\def\simpr{\lower.5ex\hbox{\prosima}}
\def\la{\lsim}
\def\ga{\gsim}

\def\specialname[#1]{\textsc{#1}}

\def\mpc{\, h^{-1}{\rm {Mpc}}}
\def\cmpc{\, h^{-1}{\rm {cMpc}}}

\def\msun{\, h^{-1}{\rm M_\odot}}

\title[]{A Conditional Abundance Matching Method of Extending Simulated Halo Merger 
Trees to Resolve Low-Mass Progenitors and Sub-halos} 

\author[Yangyao Chen et al.]{
    Yangyao Chen,$^{1,2}$\thanks{E-mail: yangyaochen.astro@foxmail.com}  
    H.J. Mo, $^{3}$   		
    Cheng Li, $^{4}$  		
    Kai Wang, $^{5}$		    
    Huiyuan Wang, $^{1,2}$ 
    and Xiaohu Yang $^{6,7}$
    \\ 
    $^{1}$School of Astronomy and Space Science, University of Science and Technology of China, Hefei, Anhui 230026, China\\
    $^{2}$Key Laboratory for Research in Galaxies and Cosmology, Department of Astronomy, University of Science and Technology of China, Hefei, Anhui 230026, China \\
    $^{3}$Department of Astronomy, University of Massachusetts, Amherst, MA 01003-9305, USA\\
    $^{4}$Department of Astronomy, Tsinghua University, Beijing 100084, China\\
    $^{5}$Kavli Institute for Astronomy and Astrophysics, Peking University, Beijing 100871, China\\
    $^{6}$Department of Astronomy, School of Physics and Astronomy, Shanghai Jiao Tong University, Shanghai, 200240, China\\
	$^{7}$Tsung-Dao Lee Institute, and Shanghai Key Laboratory for Particle Physics and Cosmology, Shanghai Jiao Tong University, Shanghai, 200240, China\\ 
}

\date{Accepted XXX. Received YYY; in original form ZZZ}

\pubyear{2021}

\begin{document}
\label{firstpage}
\pagerange{\pageref{firstpage}--\pageref{lastpage}}
\maketitle


\begin{abstract}
We present an algorithm to extend subhalo merger trees 
in a low-resolution dark-matter-only simulation by conditionally matching them 
to those in a high-resolution simulation. 
The algorithm is general and can be applied to simulation data with 
different resolutions using different target variables.
We instantiate the algorithm by a case in which trees from ELUCID, 
a constrained simulation of $(500h^{-1}{\rm Mpc})^3$ volume of the local 
universe, are extended by matching trees from TNGDark, 
a simulation with much higher resolution. Our tests show that the extended 
trees are statistically equivalent to the high-resolution trees in 
the joint distribution of subhalo quantities and in important summary 
statistics relevant to modeling galaxy formation and evolution in halos. 
The extended trees preserve certain information of individual 
systems in the target simulation, including properties of resolved satellite 
subhalos, and shapes and orientations of their host halos. With the extension, subhalo merger 
trees in a cosmological scale simulation are extrapolated to a mass 
resolution comparable to that in a higher-resolution simulation carried out in  
a smaller volume, which can be used as the input for (sub)halo-based models
of galaxy formation. The source code of the algorithm, and halo merger trees
extended to a mass resolution of $\sim 2 \times 10^8 \msun$ in the entire ELUCID 
simulation, are available.
\end{abstract}

\section{Introduction}
\label{sec:intro}

In the concordant $\Lambda$-CDM cosmology, the peaks of the density field, 
known as dark matter halos, are the building blocks of large scale structures of the 
Universe. Galaxies form and evolve through gas cooling and condensation 
in the gravitational background provided by dark matter halos 
\citep[e.g., ][]{whiteCoreCondensationHeavy1978,moGalaxyFormationEvolution2010}.
Galaxies are complex ecosystems where various components, such as dark matter, gas, 
stars and black holes, interact through complicated physical processes, presenting 
interesting and yet challenging problems for modern astrophysics. 
Enormous efforts, motivated by both theory and observation, have been made 
to model galaxy formation under various assumptions. 
Perhaps the most powerful approach to study galaxy formation 
is hydrodynamic simulation, which relies on the advances in computational 
resources and aims at modeling galaxies from first principles 
\citep[e.g.,][]{
springelCosmologicalSmoothedParticle2003, springelPurSiMuove2010,
genelIntroducingIllustrisProject2014,
vogelsbergerPropertiesGalaxiesReproduced2014,
schayeEAGLEProjectSimulating2015,crainEAGLESimulationsGalaxy2015,
pillepichFirstResultsIllustrisTNG2018,springelFirstResultsIllustrisTNG2018,
nelsonFirstResultsIllustrisTNG2018,
naimanFirstResultsIllustrisTNG2018,marinacciFirstResultsIllustrisTNG2018,
daveSimbaCosmologicalSimulations2019,
nelsonIllustrisTNGSimulationsPublic2019,
pillepichFirstResultsTNG502019,
vogelsbergerCosmologicalSimulationsGalaxy2020}. 
Here physical processes for galaxy formation are simulated with a set 
of differential equations, complemented with subgrid physics to deal 
with situations of limited numerical resolution and uncertain processes 
on small scales. With careful calibrations, hydrodynamic 
simulations can successfully reproduce many statistical properties 
of the galaxy population and provide insights into physical processes 
underlying observational data.

To overcome some of the limitations of numerical simulations, particularly 
in computational costs and numerical uncertainties, a different category of 
methods, known as halo-based semi-analytical or empirical methods, have been 
proposed. These methods simplify the modeling of galaxy formation by splitting it into 
abstract layers that are assumed to be independent. 
Specifically, these methods use dark-matter-only (DMO) simulations 
\citep[e.g., ][]{springelCosmologicalSimulationCode2005, 
boylan-kolchinResolvingCosmicStructure2009,wangELUCIDEXPLORINGLOCAL2016,
fengFastPMNewScheme2016,
habibHACCSimulatingSky2016,
wangPHoToNsParallelHeterogeneous2018,
falckIndraPublicComputationallyAccessible2021,
frontiereFarpointHighResolutionCosmology2021} as input, find (sub)halos 
using some algorithms (structure/halo finders), link (sub)halos in different snapshots 
through some tree builders, and populate (sub)halos or trees with galaxies 
using empirical relations motivated by physical and observational priors.
With such an abstraction, problems in each layer can be solved independently, 
so that the complexity in modeling the full process of galaxy formation is reduced. 
There is a vast literature in each of the steps. Examples of the structure finders 
include those based on the overdensity set obtained with boundary growing and pruning 
\citep{springelSimulationsFormationEvolution2005,
boylan-kolchinResolvingCosmicStructure2009,
planellesASOHFNewAdaptive2010,
valles-perezHaloFindingProblem2022}, and those based on direct link of 
particles \citep{davisEvolutionLargescaleStructure1985,
diemandEarlySupersymmetricCold2006,
behrooziROCKSTARPHASESPACETEMPORAL2012}.
Examples of tree builders include Monte Carlo methods 
based on the extended Press-Schechter (EPS) formalism 
(\citet{somervilleHowPlantMerger1999,coleHierarchicalGalaxyFormation2000, 
parkinsonGeneratingDarkMatter2007,
somervilleSemianalyticModelCoevolution2008,
zhangConditionalMassFunctions2008}, 
see also \citet{jiangGeneratingMergerTrees2014} for a review), 
those based on linking simulated (sub)halos 
\citep[][]{springelSimulationsFormationEvolution2005,
boylan-kolchinResolvingCosmicStructure2009,
hanResolvingSubhaloesLives2012,
behrooziGravitationallyConsistentHalo2013,
jiangNbodyDarkMatter2014a}, 
and those based on post-processing and homogenizing trees produced by 
other methods \citep[][]{hellyGalaxyFormationUsing2003,jiangNbodyDarkMatter2014a}.
Examples of halo-based models include those matching galaxies and halos based on abundance 
\citep{moStructureClusteringLymanbreak1999,valeLinkingHaloMass2004,
guoHowGalaxiesPopulate2010,simhaTestingSubhaloAbundance2012}, 
clustering \citep{guoModellingGalaxyClustering2016} 
and age \citep{hearinDarkSideGalaxy2013,hearinDarkSideGalaxy2014,
mengMeasuringGalaxyAbundance2020a,wangRelatingGalaxiesDifferent2023}, 
halo occupation distributions
\citep[HODs;][]{jingSpatialCorrelationFunction1998,
berlindHaloOccupationDistribution2002,
guoRedshiftspaceClusteringSDSS2015,
guoModellingGalaxyClustering2016,
yuanAbacusHODHighlyEfficient2022,
qinHIHODHalo2022}, 
the conditional luminosity function 
\citep[CLFs;][]{yangConstrainingGalaxyFormation2003,
zandivarezLuminosityFunctionGalaxies2006,
yangGalaxyGroupsSDSS2008,
robothamVariationGalaxyLuminosity2010,
zandivarezLuminosityFunctionGalaxies2011,
mengGalaxyPopulationsGroups2022} 
and conditional color-magnitude distribution 
\citep[CCMD;][]{xuConditionalColourMagnitude2018},
empirical models based on star formation histories of galaxies 
\citep{mutchSimplestModelGalaxy2013,
luEmpiricalModelStar2014,
luStarFormationStellar2015, 
mosterEmergeEmpiricalModel2018,
behrooziUniverseMachineCorrelationGalaxy2019,
mosterGalaxyNetConnectingGalaxies2020},
and semi-analytical models (SAMs) that emphasize more on 
physical motivated prescriptions than empirical models 
\citep[][]{
whiteGalaxyFormationHierarchical1991,
kauffmannFormationEvolutionGalaxies1993,
coleRecipeGalaxyFormation1994,
somervilleSemianalyticModellingGalaxy1999,
coleHierarchicalGalaxyFormation2000,
springelSimulationsFormationEvolution2005,
kangSemianalyticalModelGalaxy2005,
somervilleSemianalyticModelCoevolution2008,
guoDwarfSpheroidalsCD2011,
somervilleGalaxyPropertiesUltraviolet2012,
adePlanck2013Results2014,
poppingEvolutionAtomicMolecular2014,
luBayesianInferencesGalaxy2014,
henriquesGalaxyFormationPlanck2015,
laceyUnifiedMultiwavelengthModel2016,
stevensBuildingDiscStructure2016,
baughGalaxyFormationPlanck2019,
yungSemianalyticForecastsJWST2019,
henriquesLGALAXIES2020Spatially2020,
somervilleMockLightconesTheory2021,
yungSemianalyticForecastsRoman2022}.

The halo-based models described above capitalize heavily on structures 
resolved by DMO simulations. Because of computational limitations, 
these simulations always need to trade off between large simulation 
volumes and high numerical resolutions, because large volumes are 
needed to suppress cosmic variances 
\citep[e.g., ][]{somervilleCosmicVarianceGreat2004, 
mosterCOSMICVARIANCECOOKBOOK2011, chenELUCIDVICosmic2019}, 
while high resolutions are required to follow galaxy formation 
and evolution in halos/subhalos accurately. 
In particular, the properties of subhalos may not be properly 
resolved at high-$z$ when their masses are below the resolution 
limit of a large-box simulation. The limited resolution also 
makes the treatment of the evolution of satellite subhalos uncertain, as they may 
artificially lose particles and get destroyed as a result 
\citep[e.g.,][]{boschDisruptionDarkMatter2018,vandenboschDarkMatterSubstructure2018,
greenTidalEvolutionDark2021}.  
Thus, the application of a halo-based model to a cosmological-scale DMO 
simulation cannot rely solely on the assembly histories of subhalos provided 
by the simulation. Because of this, various methods have been adopted 
to extend the subhalo population in large-box simulations 
so as to trace the progenitors and subhalos that are missed.
For example, \citet{chenELUCIDVICosmic2019} used Monte Carlo trees 
generated from the EPS formalism to extend simulated 
trees in ELUCID. \citet[][]{yungSemianalyticForecastsRoman2022,yungSemianalyticForecastsJWST2022}
used EPS-based trees to replace the full assembly histories of 
halos in their adopted simulations. \citet{chenMAHGICModelAdapter2021} 
adopted the assembly histories of halos from a high-resolution DMO 
simulation to amend halo histories in a low-resolution DMO simulation, 
and found that this method is more accurate than the EPS-based amendment. 

Some efforts have been made to use satellite subhalos in simulations to 
model satellite galaxies, but many of them rely on simple assumptions.
For example,   
\citet{chenELUCIDVICosmic2019, yungSemianalyticForecastsRoman2022,
yungSemianalyticForecastsJWST2022} did not use any information  
carried by satellite subhalos in simulations. Instead, they adopted   
a dynamic friction model to predict the lifetimes of satellite subhalos/galaxies, 
and used the Navarro-Frenk-White \citep[NFW;][]{navarroUniversalDensityProfile1997} 
profiles of the host halos 
to assign phase-space coordinates (positions and velocities) to satellites. 
Because the assignment of phase-space coordinates is random and based 
on host halos in the current snapshot, the correlation of 
phase-space coordinates with other current and historical (sub)halo properties 
is lost. Consequently, the spatial distribution obtained this 
way may be biased for galaxies selected according to properties 
that are correlated to the history and environment of subhalos. 
\citet{guoRedshiftspaceClusteringSDSS2015,
guoModellingGalaxyClustering2016, yuanCanAssemblyBias2020, 
yuanAbacusHODHighlyEfficient2022,
yuanFullForwardModel2022} assigned galaxies obtained from 
HOD models to random particles in simulated halos.
As tested by \citet{boseRevealingGalaxyhaloConnection2019} with a 
hydrodynamic simulation, radial distributions of satellite 
galaxies of given stellar mass match accurately the best-fit NFW 
profiles of their host halos, which provides supports to the 
particle-based assignment scheme. However, the correlation between 
phase-space properties and other (sub)halo properties are still lost
in this scheme. 
\citet{liAIassistedSuperresolutionCosmological2021,
niAIassistedSuperresolutionCosmological2021} extended low-resolution
DMO simulations by populating more particles in the simulation volumes, 
using deep learning models trained by high-resolution simulations.
This method preserves environmental information of the 
low-resolution simulation, but again, the extension is made at separate 
snapshots and thus loses information about subhalo formation histories.
The two semi-analytical models of 
GALFORM \citep{coleHierarchicalGalaxyFormation2000,
laceyUnifiedMultiwavelengthModel2016,
baughGalaxyFormationPlanck2019} 
and L-Galaxies \citep{henriquesGalaxyFormationPlanck2015,
henriquesLGALAXIES2020Spatially2020} 
used simulated phase-space information of satellite subhalos before 
they are disrupted, 
and linked a modeled ``orphan'' galaxy, whose subhalo has been artificially 
disrupted, to the most bound particle of its subhalo just before disruption.
This choice preserves some of the correlations of subhalos
described above, but may introduce some other problems.  
For example,  the most bound particles may be biased tracers of their 
subhalos after disruption, and a single particle in a shallow 
potential may accidentally lose its binding energy and jump to 
an unrelated location owing to numerical effects. 
Perhaps the ultimate solution to reliably resolving satellite 
subhalos is to use zoom-in simulations of individual sub-regions of interest 
\citep[e.g.,][]{kangSemianalyticalModelGalaxy2005,
barnesClusterEAGLEProjectGlobal2017,nelsonIllustrisTNGSimulationsPublic2019}.
However, such high-resolution zoom-in simulations are still computationally 
expensive and thus infeasible to cover the volume of a large cosmological simulation.

To build a solid foundation for halo-based models, we develop in this paper 
a powerful algorithm to extend the resolution of subhalo merger trees in a 
low-resolution DMO simulation by conditionally matching them with those in 
another high-resolution DMO simulation. The extended trees have more 
complete assembly histories for low-mass halos at high-$z$, 
and satellite subhalos extend their lifetimes with assigned phase-space 
coordinates after they are disrupted by numerical effects.  
As we will show, the extension algorithm not only reproduces the joint 
distribution of various subhalo properties, including their phase-space coordinates, 
but also tries to maximally keep information about individual systems
resolved by the target low-resolution simulation, such as properties of 
satellite subhalos and shapes of their host halos.
With such an extension, halo-based galaxy formation models 
can be built on more complete (sub)halo assembly histories and 
more reliable predictions for the galaxy population.  

This paper is organized as follows. In \S\ref{sec:data}, we introduce the 
simulation data used in our analysis. In \S\ref{sec:method}, we describe 
the algorithm to extend subhalo merger trees. We first present a general 
scheme that is applicable to a wide range of input data, and then specify 
cases studied in the present paper. In \S\ref{sec:results}, we present 
tests on the performance of the extension on various properties of the 
merger trees and the subhalo population. Finally, we summary and discuss 
our main results in \S\ref{sec:summary}. Code and data availability 
are described in the end of the main text.


\section{Simulation Data}
\label{sec:data}

\begin{center}
\begin{table*} 
\caption{Cosmologies and simulation parameters of simulations used in this paper.
    Box size $L_{\rm box}$, 
    number of resolution units $N_{\rm resolution}$, 
    dark matter particle mass $m_{\rm dark\ matter}$,
    and target baryon mass {\revstyle resolution} $m_{\rm baryon}$ are listed in 
    different columns. 
    $N_{\rm resolution}$ in TNG is the total number of dark matter particles 
    and the initial number of gas cells. 
    $N_{\rm resolution}$ in TNGDark and ELUCID 
    is the number of dark matter particles. }
\begin{tabularx}{\textwidth}{c | X | >{\hsize=.15\hsize}X | >{\hsize=.15\hsize}X >{\hsize=.15\hsize}X >{\hsize=.13\hsize}X}
    \hline 		
        {\bf Simulation}   			
        & {\bf Cosmology}
        & $L_{\rm box} $  \newline $[\cmpc]$
        & $N_{\rm resolution}$    
        & $m_{\rm dark\ matter}$   \newline $[\msun]$
        & $m_{\rm baryon}$			\newline $[\msun]$
        \\
    \hline
        TNG 					
        & \multirow{2}{9.0cm}{
            Planck15 \citep{adePlanck2015Results2016}: 
            $h=0.6774$, $\Omega_{\Lambda,0}=0.6911$, $\Omega_{M,0}=0.3089$, 
            $\Omega_{B,0}=0.0486$, $\Omega_{K,0}=0$, $\sigma_8=0.8159$, $n_s=0.9667$}
        & \multirow{2}{9.0cm}{75}
        & $2 \times 1820^3$
        & $5.1\times 10^6$ 
        & $9.4\times 10^5$
        \\
    \cline{1-1}\cline{4-6}
        TNGDark 	
        &
        & 
        & $1820^3$
        & $6.0\times 10^6$ 
        & -
        \\ & & & & & \\
    \hline
        ELUCID					
        & WMAP5 \citep{dunkleyFIVEYEARWILKINSONMICROWAVE2009}: 
            $h=0.72$, $\Omega_{\Lambda,0}=0.742$, 
            $\Omega_{M,0}=0.258$, $\Omega_{B,0}=0.044$, $\Omega_{K,0}=0$, 
            $\sigma_8=0.80$, $n_s=0.96$
        & $500$
        & $3072^3$
        & $3.08 \times 10^8$
        & -	
        \\
    \hline
\end{tabularx}
\label{tab:simulations}
\end{table*}
\end{center}

Throughout this paper, we use two N-body simulations to implement and test 
the extension of subhalo merger trees. 

The first is ELUCID \citep{wangELUCIDEXPLORINGLOCAL2016}, a DMO
simulation obtained using the N-body code \specialname[L-Gadget], 
a memory optimized 
version of \specialname[Gadget-2] \citep{springelCosmologicalSimulationCode2005}. 
A total of 100 snapshots, from redshift $z=18.4$
to $0$, are saved. 
Halos are identified with the friends-of-friends (\specialname[FoF]) 
algorithm \citep{davisEvolutionLargescaleStructure1985} with a scaled linking length of $0.2$. 
Subhalos are identified with the \specialname[Subfind] 
algorithm \citep{springelPopulatingClusterGalaxies2001,dolagSubstructuresHydrodynamicalCluster2009}, 
and subhalo merger trees are constructed using the \specialname[SubLink]
algorithm \citep{springelCosmologicalSimulationCode2005,boylan-kolchinResolvingCosmicStructure2009}.
ELUCID has a simulation box with side length of $500\mpc$ and uses
a total of $3072^3$ particles to trace the cosmic density field. The mass of 
each dark matter particle is $3.08\times 10^8 \msun$ and the mass resolution limit of 
FoF halos is about $10^{10} \msun$.

The second simulation is TNG100-1-Dark, a run of the Illustris-TNG project
\citep{nelsonIllustrisTNGSimulationsPublic2019,pillepichFirstResultsIllustrisTNG2018,
springelFirstResultsIllustrisTNG2018,nelsonFirstResultsIllustrisTNG2018,
naimanFirstResultsIllustrisTNG2018,marinacciFirstResultsIllustrisTNG2018},
which is a suite of cosmological hydrodynamic simulations carried out with 
the moving mesh code Arepo \citep{springelPurSiMuove2010}. 
Processes for galaxy formation, such as gas cooling, star formation, 
stellar feedback, metal enrichment, and AGN feedback, 
are simulated with subgrid prescriptions tuned to match a set of observational data
\citep[see][]{weinbergerSimulatingGalaxyFormation2017,
pillepichSimulatingGalaxyFormation2018}. 
A total of 100 snapshots, from redshift $z=20.0$ to $0$, are saved for 
each run. Halos, subhalos and subhalo merger trees are identified 
and constructed using the same algorithms as ELUCID, with modifications to
include stellar particles and gas cells in the identification of subhalos 
\citep[see, e.g.,][for a summary]{rodriguez-gomezMergerRateGalaxies2015}. 
Here, we choose the TNG100-1-Dark run, the DMO counterpart of 
the full hydro run, TNG100-1. TNG100-1-Dark (thereafter TNGDark) has a 
simulation box with side length of $75 \mpc$.
The mass of each dark matter particle is $6\times 10^6 \msun$ and 
the mass resolution of FoF halos is about $2\times 10^8 \msun$.

The usage of two simulations with different cosmologies is a deliberate 
choice to test their effects on the extended subhalo merger trees.
{\revstyle
In real applications, the cosmology of the low-resolution simulation 
should exactly match that of the high-resolution simulation.}
To also test effects of baryonic processes on subhalo merger trees,
we use the TNG100-1 run (thereafter TNG) in some of our analyses.
Cosmological and simulation parameters of all the three simulations 
are listed in Table~\ref{tab:simulations}.

\section{The Extension Algorithm}
\label{sec:method}

As shown in \citet{chenELUCIDVICosmic2019,chenMAHGICModelAdapter2021}, 
subhalo merger trees in a low-resolution simulation like ELUCID are not 
sufficiently complete to use directly in empirical models of galaxy formation.
This incompleteness comes in two different ways in the evolution
history of a typical subhalo:
\benum
\item 
For a central subhalo that is resolved by the simulation at some 
redshift, part of its assembly history may be missed at higher redshift 
when its mass goes below the resolution limit.
\item 
After a subhalo falls into its host halo, the simulation may not be able to 
trace it reliably because of strong environmental effects that are 
not well modeled by the simulation.  As a result, the motion of the 
subhalo may not be well traced, and the subhalo may be disrupted 
artificially \citep[see, e.g., ][]{boschDisruptionDarkMatter2018,
vandenboschDarkMatterSubstructure2018,greenTidalEvolutionDark2021}.
\eenum
Note that such incompleteness affects not only low-mass subhalos, 
but also massive ones because massive subhalos have low-mass progenitors 
at high-$z$. To tackle the problem of limited resolution in large-box 
simulations, some expedient methods have been adopted to amend the 
simulated merger trees statistically.
For example, \citet{chenELUCIDVICosmic2019} planted small seeds of 
galaxies in central subhalos when they first became resolved in the simulation.  
\citet{luEmpiricalModelStar2014,luGalaxyEcosystemsGas2015,
chenELUCIDVICosmic2019,yungSemianalyticForecastsRoman2022,
yungSemianalyticForecastsJWST2022} 
deliberately avoided using properties of simulated subhalos after 
they are accreted by their hosts, but assigned random positions and 
velocities to these subhalos according to some assumed density 
profiles.

Here, we develop a new algorithm to extend the resolution limit of subhalo merger 
trees. The key of this algorithm is to learn tree properties from a high-resolution
simulation first, and then to extend trees in the target, lower-resolution  
DMO simulation by conditionally matching subhalos between the two simulations.
This algorithm has the following advantages:  
(i) subhalo evolution histories at high-$z$ and after infall are both complete
in the amended trees; 
(ii) distribution of subhalo properties in the high-resolution simulation are 
retained in the amended trees;
(iii) subhalo properties in the target simulation are retained as long as they 
are resolved by target simulation; 
(iv) host halo properties in the target simulation, such as shape and orientation, 
are preserved.
The extended trees thus provide a solid foundation to construct  halo-based 
models of galaxy formation. 

\begin{figure} \centering
    \includegraphics[width=0.75\columnwidth]{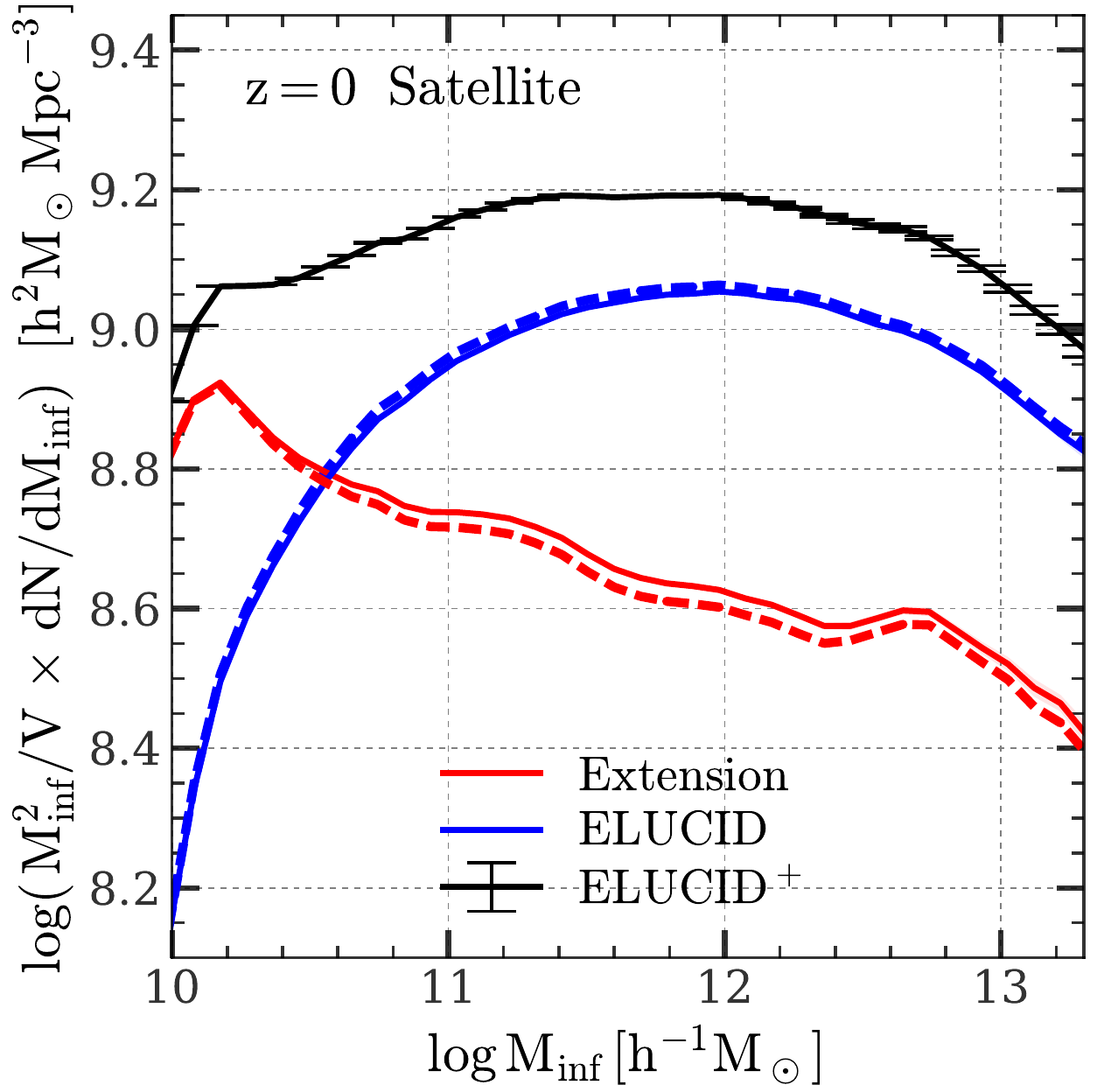}
    \caption{Infall mass functions of satellite subhalos selected at $z=0$ in 
        the ELUCID simulation. The blue solid line (labeled ``ELUCID'') is the 
        result using subhalos resolved by the original ELUCID simulation. 
        The black solid line (labeled ``ELUCID$+$'') is the result obtained 
        from amended merger trees. For reference, the red solid line 
        (labeled ``Extension'') is the result for subhalos generated by the 
        extension algorithm. A small fraction of the resolved subhalos in 
        ELUCID is moved to ``Extension'' to ensure a consistent halo-centric 
        radial distribution with the high-resolution simulation, TNGDark, 
        and the amount is the difference between the dash line (before the move) 
        and the solid line (after the move). 
        See \S\ref{ssec:specific-elucid-and-tngdark} for a detailed description. 
        The mass functions are multiplied by $M^2_{\rm inf}$ for clarity. Error 
        bars and shaded areas indicate the standard deviations computed from
        {\revstyle 50 bootstrap resamplings over halos,
        }which are too small to see owing to the large sample size of ELUCID.
    }
    \label{fig:extension-hmf}
\end{figure}

As a demonstration of the effect of extending subhalo merger trees, 
Fig.~\ref{fig:extension-hmf} shows the mass function of subhalos 
at the time of infall. Throughout this paper, we use the ``top-hat''
mass of the host FoF of a subhalo. This halo mass is calculated within 
a virial radius within which the mean density is equal to that given
by the spherical collapse model \citep{bryanStatisticalPropertiesXRay1998}. 
As our convention, we use ELUCID to denote 
the results obtained from the original ELUCID data, and $\rm ELUCID^+$ to 
denote the results obtained from amended subhalo merger trees.   
In the figure, the results obtained from ELUCID and $\rm ELUCID^+$
are shown by the solid blue and solid black lines, respectively. 
For reference, the red solid curve, marked as ``Extension'', is the 
mass function of subhalos produced by the extension algorithm.  
Comparing the simulated and amended mass functions, one can see that
the extension has a moderate effect, $\approx 0.15\ {\rm dex}$, 
at the high-mass end ($M_{\rm inf} > 10^{11.5} \msun$), and becomes more significant 
for subhalos of lower mass, reaching to more than $0.6\ {\rm dex}$ 
at the lowest-mass end ($M_{\rm inf} = 10^{10} \msun$).
Because low-mass systems dominate the subhalo population, 
amended summary statistics of subhalos are expected to be significantly 
different from those derived from the original simulation, indicating the 
importance of the amendment in modeling the subhalo population reliably.  

For brevity, we only show the results for subhalos at $z=0$ in the main text 
to demonstrate the performance of our extension algorithm. 
Our tests showed that the extension algorithm actually works as well at high-$z$, 
because the density field is less evolved and the halo population 
is less diverse (see Appendix~\ref{app:high-z} for the details).

\begin{figure*} \centering
    \includegraphics[width=0.68\textwidth]{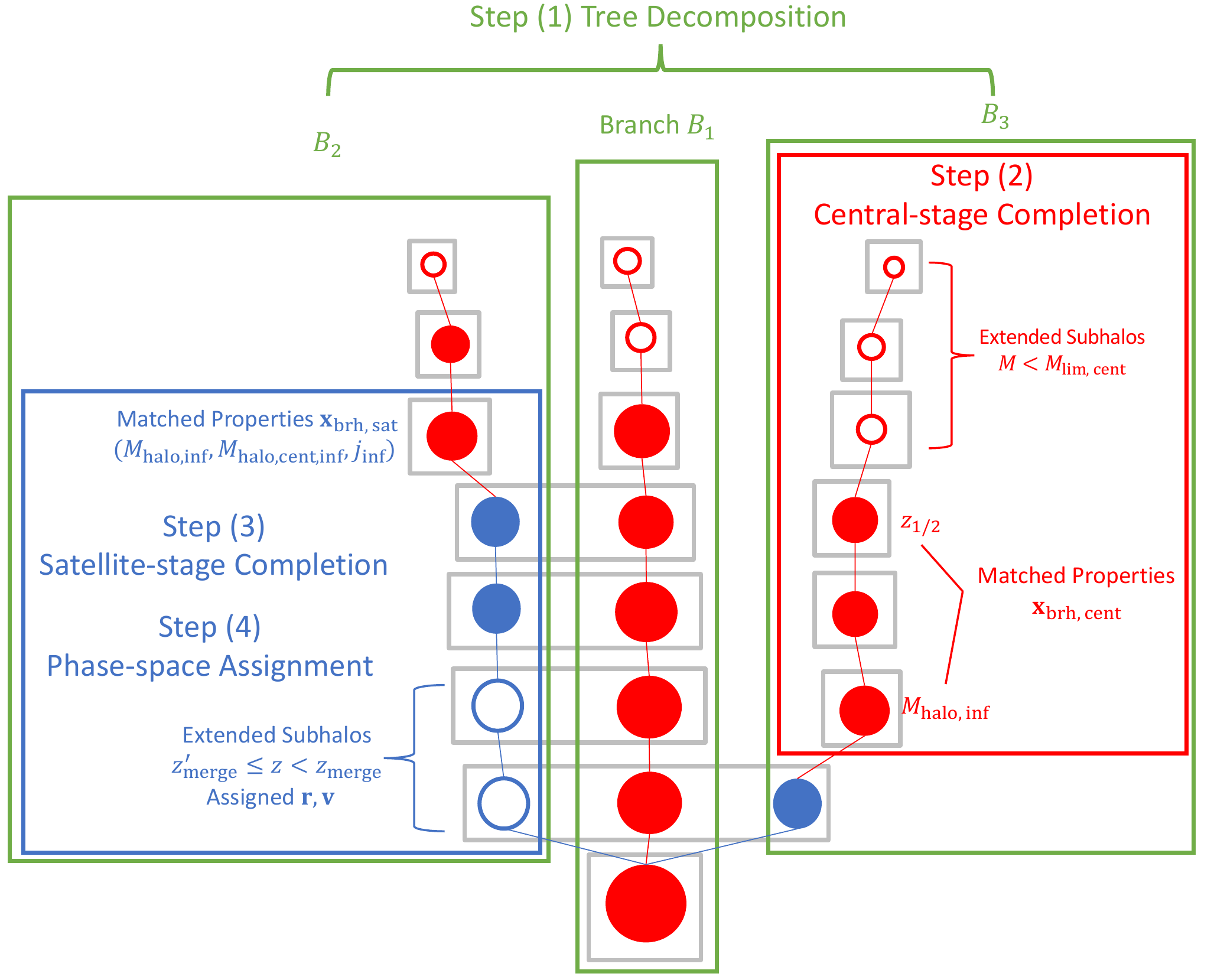}
    \caption{\revstyle
        A schematic diagram of the subhalo merger tree extension algorithm, 
        as described in Table~\ref{tab:notations-in-algorithm} and 
        elaborated upon in \S\ref{ssec:extension-algorithm}.
        Gray boxes represent halos,
        with red and blue circles representing central and satellite subhalos, 
        respectively. 
        Filled circles denote subhalos that are resolved by the simulation, 
        while empty circles indicate subhalos that were missed and subsequently 
        created through the extension.
        Subhalos processed at each step of the algorithm are enclosed within 
        a colored box.
    }
    \label{fig:schematic}
\end{figure*}

The rest of this section is organized as follows. 
In \S\ref{ssec:outline-algorithm}, we outline the algorithm by 
listing its four steps.
In \S\ref{ssec:extension-algorithm}, we describe each of the steps
in general terms, so that the algorithm can be adapted to  
different target variables and to subhalo merger trees with 
different resolutions. In \S\ref{ssec:specific-elucid-and-tngdark}
we describe the application of the general framework to a specific case 
of amending subhalo merger trees of ELUCID with the use of TNGDark.
For reference, Table~\ref{tab:notations-in-algorithm} summarizes  
the notations of variables to be used in the description of the 
general framework, and Table~\ref{tab:specific-choices}
summarizes the notations in the description of the specific case
of using TNGDark to amend ELUCID merger trees. {\revstyle
Fig.~\ref{fig:schematic} shows a schematic diagram of the algorithm.
}

\begin{table*}
\caption{Notations for variables used in the description of the extension 
algorithm in \S\ref{ssec:extension-algorithm}. 
The first column lists the location where the notation first appears.
The second and third columns list the notations and their descriptions, 
respectively.
Note that most of these are abstract variables used in the description of 
the general framework. The concrete choices depend on the specific application
(see \S\ref{ssec:specific-elucid-and-tngdark} and Table~\ref{tab:specific-choices} 
for the example demonstrated in this paper). 
}
\begin{tabular}{ p{3cm} p{3.5cm} p{10.0cm}  }
    \hline
    {\bf First Appearance} & {\bf Notations} & {\bf Descriptions} \\
    \hline
    Outline of the Algorithm   
        &  $S$, $S'$             & The target low-resolution simulation, and the reference high-resolution simulation used as training source. \\
    \hline
    \multirow{3}{3cm}{Tree decomposition}
        &  $F$, $T$              & A forest and a subhalo merger tree. \\
        &  $B_i$, $r_i$, $c_i$   & The $i$-th branch obtained by decomposing a subhalo merger tree, the root subhalo of this branch, and the ``last central subhalo'' of this branch. \\
        &  $N_B$                 & The number of branches obtained by decomposing a subhalo merger tree. \\
        &  $z_{\rm inf}$, $z_{\rm first}$ &  The infall redshift of a whole branch or of any subhalo in this branch, and the first resolvable redshift of this branch. \\
    \hline
    \multirow{3}{3cm}{Central-stage completion}
        & ${\bf x}_{\rm brh, cent}$ & A set of branch properties used to match central stages of branches.  \\
        & $d_{\rm cent}(B, B')$     & The $L_2$ distance between two branches $B$ and $B'$ for the central stage. \\
        & $M_{\rm lim, cent}$, $z_{\rm joint}$ & The halo mass threshold below which the extension is applied for a branch, and the corresponding ``joint'' redshift. \\
    \hline
    \multirow{3}{3cm}{Satellite-stage completion}
        & ${\bf x}_{\rm brh, sat}$ & A set of branch properties used to match satellite stages of branches. \\
        & $d_{\rm sat}(B, B')$ & The $L_2$ distance between two branches $B$ and $B'$ for the satellite stage. \\
        & $z_{\rm merge}$  & The redshift when a satellite subhalo merges into another subhalo. \\
    \hline
    \multirow{7}{3cm}{Phase-space assignment}
        & ${\bf x}_{\rm sat}$ & The set of satellite properties whose joint distribution is required to be recovered when we assign properties to satellites. \\
        & ${\bf x}_{\rm sat, complete}$, ${\bf x}_{\rm sat, incomplete}$ & The complete and incomplete parts of ${\bf x}_{\rm sat}$ that are resolved and missed by the target simulation, respectively. \\
        & $I_{\rm missed}$ & A binary variable indicating whether or not a satellite is missed by the target simulation. \\
        & $C_i$, $H_i$, $N_{H_i}$ & The $i$-th cell obtained by partitioning the feature space of satellites, the set of satellite subhalos in this cell, and the size of this set. \\
        & $d_{\rm cell}(H_i, H_j')$ & The $L_2$ distance between two cells $H_i$ and $H_j'$ in the match of conditioning variables. \\
        & $ N_{\rm cell} $, $N_{\rm cell, max}$,  & The total number of cells and its upper bound imposed by us. \\
        & $N_{\rm min, cell\ partition}$, $N_{\rm min, cell\ match}$ &  The minimal number of satellites from $S$ and $S'$, respectively, for a cell to be treated as valid.  \\
    \hline
\label{tab:notations-in-algorithm}
\end{tabular}
\end{table*}

\begin{table*}
    \caption{
        Summary of notations (first panel) and choices (second panel) specific 
        to $S={\rm ELUCID}$ and $S'={\rm TNGDark}$ used in \S\ref{ssec:specific-elucid-and-tngdark}. 
        Some intermediate variables are not listed here. 
        A variable that appears in both $S$ and $S'$ is distinguished by 
        a prime symbol, such as $r_{\rm lf}$ and $r_{\rm lf}'$.
    }
    \begin{tabular}{ p{3cm} p{12cm}  }
    \hline
    {\bf Notations} & {\bf Descriptions} \\
    \hline
    $M_{\rm halo,inf}$ & The infall mass of a whole branch or of any subhalo (central or satellite) in this branch. \\
    $M_{\rm halo, host}$ & The mass of the current host halo of any subhalo (central or satellite). \\
    $M_{\rm inf,sat}$, $M_{\rm halo, cent, inf}$, $j_{\rm inf}$   & 
        For any satellite subhalo, these three variables are the halo mass of it right before infall,
        the halo mass of the central subhalo into which it falls, 
        and its orbital angular momentum, respectively. \\
    $M_{\rm match, cent}$ & The threshold of $M_{\rm halo, inf}$ below which formation time is not used for the central-stage neighbor matching. \\
    $z_{\rm 1/2}$  & The half-halo-mass formation redshift of a central subhalo, i.e., the redshift at which the halo mass on 
        its main branch first exceeds half of its current halo mass. \\

    & \\
    ${\bf r}_{\rm p, i}$, ${\bf v}_{\rm p, i}$, $N_{\rm p}$ &
        The position and velocity of the $i$-th particle in a halo, and the total number of particles in that halo. \\
    $\mathcal{I}$, $\lambda_i$, ${\bf e}_i$, $a_i$, $s_i$  &
        For a halo, these give its inertial tensor, 
        the $i$-th eigenvalue and eigenvector of the inertial tensor,
        the $i$-th major axis of the inertial ellipsoid,  
        and the stretching factor along this axis, respectively (see Eqs.~\ref{eq:inertial-tensor}, \ref{eq:eigen-axes} and \ref{eq:stretch-factor}). \\
    ${\bf r}_{\rm com}$, ${\bf v}_{\rm com}$ &
        The position and velocity of the center of mass (COM) of a halo. \\
    $R_{\rm halo, host}$, $R_{\rm halo, host}$ &
        The virial radius and virial velocity of the host halo of a subhalo.  \\
    
    & \\
    ${\bf r}$, ${\bf v}$ & The position and velocity of a subhalo in real space. \\
    ${\bf r}_{\rm lf}$, ${\bf v}_{\rm lf}$ & The position and velocity of a subhalo in the local frame defined by its host halo (see Eq.~\ref{eq:local-frame-transformation}). \\    
    $r_{\rm lf}$, $\theta_{r, {\rm lf}}$, $\phi_{r, {\rm lf}}$ &
        The spherical coordinates of the local-frame position. \\
    $v_{\rm lf}$, $\theta_{v, {\rm lf}}$, $\phi_{v, {\rm lf}}$ &
        The spherical coordinates of the local-frame velocity. \\
    $r_{\rm lf, com}$ & 
        For a halo, this variable gives the distance between its COM
        and the minimal potential of its central subhalo, both measured in 
        the local frame. This variable is an
        indicator to the relaxation state of a halo. \\
    $\Delta \log r_{\rm lf, max}$ & The maximal difference in the halo-centric distance
        for a subhalo in $S$ to be conditionally matched with a subhalo in $S'$. \\
    \hline
    \vspace{0.5cm}
    \end{tabular}
    \begin{tabular}{ p{3cm} p{7.5cm}  }
        \hline
        {\bf Step } & {\bf Choices} \\
        \hline
        Central-stage completion
            & ${\bf x}_{\rm brh, cent} = \,[\log M_{\rm halo,inf},\, \log(1+ z_{1/2}) \,]$ or $\log M_{\rm halo,inf}$ \\
            & $M_{\rm match,cent}=2\times 10^{10} \msun$, $M_{\rm lim,cent}=10^{10} \msun$ \\
        & \\
        Satellite-stage completion 
            & ${\bf x}_{\rm brh, sat} = (\log M_{\rm halo, inf},\,\log M_{\rm halo, cent, inf}, \log\,j_{\rm inf})$  \\
        & \\
        Phase-space assignment
            & $N_{\rm cell, max}$=768, $N_{\rm min, cell\ partition} = 32$, $N_{\rm min, cell\ match}=32$ \\
            & ${\bf x}_{\rm sat, complete} = [\log (1+z_{\rm inf}), \log \frac{M_{\rm inf, sat}}{M_{\rm halo, host}}, \log M_{\rm halo, host}, \ r_{\rm lf, com} ]$ \\        
            & ${\bf x}_{\rm sat, incomplete} =  \, ({\bf r}_{\rm lf},\,{\bf v}_{\rm lf})$ \\
            & $\Delta \log r_{\rm lf, max}$ = 0.1 \\
        \hline
    \end{tabular}
\label{tab:specific-choices}
\end{table*}

\subsection{Outline of the Algorithm}
\label{ssec:outline-algorithm}

The extension algorithm is designed to work on all subhalo merger trees in a low-resolution 
simulation, $\rm S$, by learning from another high-resolution simulation, $\rm S'$.
The goal is that, for any central subhalo identified in $\rm S$, 
(i) its mass assembly history is extended to higher redshift with 
a mass resolution similar to that of $\rm S'$, 
and (ii) its lifetime after infall is extended to be consistent 
with that expected from $\rm S'$. Note that we cannot create a 
subhalo whose mass is always below the resolution limit of $S$, 
so that it is not identifiable in $S$. 
{
\revstyle
In Appendix~\ref{app:ssec:completeness}, we examine the completeness of the 
extended population and the effects of these completely missed branches.
}
The algorithm consists of the following main steps: 
\benum
\item 
Tree decomposition: 
each subhalo merger tree in $S$ or $S'$ is 
decomposed into disjoint branches. These branches will be 
used as pieces to complete trees of subhalos in both 
central and satellite stages described in the following two steps.   
\item 
Central-stage completion: 
the mass assembly history (MAH) of any central subhalo, defined as the 
set of halo mass values in the main branch of the subhalo merger tree 
rooted in this subhalo, is completed down to the same mass limit as $\rm S'$. 
With this step, the mass assembly histories of all central subhalos in $S$ 
are extended well below the mass limit of $\rm S$, so that 
empirical models applied to them can trace star formation 
in a galaxy to high redshift when the amount of stars formed in galaxy 
is insignificant. This step is decoupled from the next two 
steps, so that it can be skipped if the MAH of a central subhalo 
does not need to be extended.
\item 
Satellite-stage completion: 
the lifetime of a subhalo in $S$ after the infall is extended
so that it is not artificially destroyed due to the limited resolution of $S$.
The links of subhalos in merger trees are updated to reflect the addition 
of subhalos generated by the extension.
With this step, the number of satellite subhalos in a host halo 
is similar to that expected in the high-resolution simulation.
Thus, empirical models applied to $S$ will be able to describe
the satellite population conditioned on host halos, such as the 
conditional galaxy stellar mass functions (CGSMFs), satellite density profiles, 
and the one-halo terms of two-point correlation functions (TPCFs). 
\item 
Assignment of phase-space coordinates to satellite subhalos: 
positions and velocities are assigned to all the satellite subhalos,
both the original population and the population generated by the 
extension algorithm. In this step, subhalo properties, such as spatial 
position, velocity, and various properties at the time of infall, 
are required to be statistically recovered.
Phase-space properties of satellite subhalos that are resolvable 
by $\rm S$ are kept unchanged whenever possible.
Properties of host halos, such as their shapes and orientations, 
are also preserved whenever possible. With this strategy, the 
algorithm retains all reliable information from the original 
simulation, and perform extensions only when necessary.
\eenum

\subsection{Details of the Algorithm}
\label{ssec:extension-algorithm}

\subsubsection{Tree Decomposition} 
\label{sssec:tree-decomposition}

In the tree decomposition step, we aim to split each subhalo merger 
tree, $T$, into a set of disjoint branches $\{ B_i \} _{i=1} ^ {N_B}$, 
each consisting of a chain of subhalos that form the main branch of a root 
subhalo, $r_i \in B_i$. Here, $N_B$ is the number of branches in $T$,
and $\cup_{i=1}^{N_B} B_i = T$. 
{
The decomposition starts from a forest
$F = \{ T \}$ that initially contains only the target tree $T$, and proceeds 
through the following substeps:
\benum
\item We arbitrarily take a tree, $T_i \in F$, out of the forest $F$, and 
we denote the root subhalo of $T_i$ as $r_i$.
\item We extract the main branch, $B_i$, of $r_i$, out of $T_i$, and we add 
$B_i$ into the result set of branches.
\item The remaining subhalos in $T_i$ form a set of sub-trees of $T_i$. 
We add all these sub-trees back into $F$. 
\item We go back to the first substep and proceed iteratively until $F$ 
becomes empty.
\eenum
}
For each branch, $B_i$, we walk through it from the root subhalo, $r_i$, 
towards high redshift, until we encounter a central subhalo $c_i \in B_i$.
We refer to this central subhalo as the ``last central subhalo'' of this branch,
and define its redshift to be the infall redshift, $z_{\rm inf}$, of the 
whole branch, and of any subhalo in this branch. 
Other properties of the last central subhalo, such as its halo mass, 
the mass of the target halo into which it is merging, 
and its orbital angular momentum relative to the target halo, 
are all computed and defined as the infall properties of the whole branch 
and of any subhalo in this branch. 
{
We refer to the subhalo with the highest redshift on $B_i$ as the ``first resolvable 
subhalo'' of this branch, and we define its redshift to be the 
first resolvable redshift, $z_{\rm first}$, of this branch.
}

\subsubsection{Central-stage Completion} 
\label{sssec:central-stage-completion}

In the central-stage completion step, we only focus on the central 
part, which consists of subhalos at or before $z_{\rm inf}$ of each branch.
For each target branch $B$ in the low-resolution simulation $S$, 
we search a reference branch $B'$ with the same infall redshift in the 
high-resolution simulation $S'$. We require that $B$ be closest to $B'$ according 
to some matching (``distance'') criteria (to be specified below). Such match allows 
subhalo properties in the history of $B'$ to be borrowed by its nearest neighbor $B$ for 
extensions of properties that are poorly resolved in $S$. 
This method, referred to as the nearest neighbor matching (NNM) in the following, 
is, effectively, a k-nearest neighbors (kNN) regression with $k = 1$,
a non-parametric regression capable of dealing with 
highly non-linear patterns in feature space of any dimensionality 
\citep[e.g.,][]{bishopPatternRecognitionMachine2006,jamesIntroductionStatisticalLearning2013}.
The general requirement of kNN is that the distributions of properties 
to be matched are similar in the two datasets. In our NNM, this 
requirement is achieved by using only properties that are 
robustly determined in both $S$ and $S'$, and by standardizing 
these properties before the matching (see \S\ref{ssec:specific-elucid-and-tngdark}).
Based on these considerations, the match between $B$ and $B'$, the truncation 
of $B$ and the borrowing from $B'$ by $B$ will be achieved through the following substeps:
\benum
\item 
We define a set of branch properties that can be reliably resolved for any branch
in both $S$ and $S'$. We denote these properties collectively as 
${\bf x}_{\rm brh, cent}$ and ${\bf x}_{\rm brh, cent}'$ in the two 
simulations, respectively. The branch properties to use should include variables 
that are the most relevant to the MAH of the central part in a branch.
\item 
For each branch $B$ in $S$, we search among all branches of the same $z_{\rm inf}$ 
in $S'$ to find a $B'$ that is closest to $B$. Here, the distance, $d_{\rm cent}(B, B')$,
between two branches, is the $L_2$ distance between ${\bf x}_{\rm brh, cent}$
and ${\bf x}_{\rm brh, cent}'$, defined as 
\begin{equation}
\begin{split}
    d_{\rm cent}(B, B') 
    & = \|
        {\bf x}_{\rm brh, cent} - {\bf x}_{\rm brh, cent}' \| \\ 
    & = \sqrt{({\bf x}_{\rm brh, cent} - {\bf x}_{\rm brh, cent}')^2}.
\end{split}
\end{equation}
\item 
The MAH of $B$ before a joint redshift, $z_{\rm joint}$, when its mass 
goes below the resolution limit, $M_{\rm lim, cent}$, is truncated 
and replaced with the MAH of $B'$ at $z>z_{\rm joint}$.
Note that the MAH of $B'$ is re-scaled  to avoid any discontinuity
around the joint redshift. Because of the difference in redshift 
sampling between $S$ and $S'$, we linearly interpolate the MAH of $B'$ 
to the redshift needed by $B$. The re-scaling and interpolation are
in logarithmic scale for MAH and in $\log(1+z)$ for the redshift.
{
After this substep, the MAH of $B$ is extended from $z_{\rm first}$ to 
the first resolvable redshift, $z_{\rm first}'$, of $B'$.
}
\item 
A list of new central subhalos, whose halo masses are defined by 
the extended part of MAH, are created and attached to the tree. 
{
To be maximally compatible with $S$, the positions and peculiar velocities of 
these subhalos at $z_{\rm first} \geqslant z > z_{\rm joint}$ in the extension
retain their simulated values in $S$.
For the sake of completeness, the peculiar velocities of these
subhalos at $z_{\rm first}' \geqslant z > z_{\rm first}$ in the extension 
are all assigned to be zero, and their spatial 
positions are set to the simulated position of the subhalo 
at $z_{\rm first}$ on $B$. 
}
This choice for assigning phase-space coordinates has no significance, because 
it is not used anywhere in empirical models of galaxy formation.
\eenum

By using the branches in the reference simulation $S'$, the 
extended MAHs are more precise than the method used in 
\citet{chenELUCIDVICosmic2019} and 
\citet{yungSemianalyticForecastsJWST2022,yungSemianalyticForecastsRoman2022}, 
where EPS-based Monte Carlo trees are used. This is due to the fact that different
EPS-based methods may produce statistically different trees 
\citep{jiangGeneratingMergerTrees2014}, 
and EPS-based methods need to be calibrated by N-body simulations
\citep{parkinsonGeneratingDarkMatter2007}. 
Even with such calibration, EPS-trees may not be able
to match simulated trees accurately \citep[e.g.,][]{chenELUCIDVICosmic2019}.

The extension of trees in the satellite stage is more complicated and we split 
it into two steps. The first is to extend the lifetimes of subhalos after 
the infall, and the second is to assign phase-space quantities to 
subhalos in their host halos. The complexity comes from the fact that 
satellite subhalos are subject to strong environmental effects, 
which need to be treated properly in order to correctly predict 
their properties, such as lifetimes, spatial positions and velocities.
Since phase-space properties of satellite subhalos can be 
observed, e.g., using the TPCFs
of galaxies in real and redshift space and the number density profiles 
of galaxies around halos \citep[e.g.,][]{zehaviLuminosityColorDependence2005,
liDependenceClusteringGalaxy2006,wangCrossCorrelationGalaxiesDifferent2007,
liDistributionStellarMass2009,shiMappingRealSpace2016,
coilPRIMUSMathplusDEEP22017,shiMappingRealSpace2018,
banerjeeNearestNeighborDistributions2020,
brainerdLopsidedSatelliteDistributions2020,
mengMeasuringGalaxyAbundance2020a,
martin-navarroAnisotropicSatelliteGalaxy2021a,
banerjeeCosmologicalCrosscorrelationsNearest2021}
it is necessary for our algorithm to recover them properly. 

\subsubsection{Satellite-stage Completion} 
\label{sssec:satellite-stage-completion}

In the satellite-stage completion step, we focus only on 
subhalos at and after $z_{\rm inf}$ in each branch. For each target 
branch $B$ in $S$, the procedure is similar to the NNM  
adopted in the central-stage completion: we search in $S'$ 
a reference branch $B'$ that matches $B$ the best in infall 
redshift and other properties, and we extend the lifetime of 
$B$ after infall using that of $B'$.
The details are contained in the following substeps:
\benum
\item 
We define a set of branch properties that can be reliably 
resolved for any branch in both $S$ and $S'$, and we denote it by 
${\bf x}_{\rm brh, sat}$ in $S$, 
and ${\bf x}_{\rm brh, sat}'$ in $S'$. Here, the 
set of branch properties chosen needs to be correlated with
the lifetime of a satellite subhalo before it merges into another subhalo.
\item 
For each branch $B$ in $S$, we match it to a branch $B'$ in $S'$ by requiring that
the $L_2$ distance, defined as \[
    d_{\rm sat}(B, B') = \| {\bf x}_{\rm brh, sat} - {\bf x}_{\rm brh, sat}' \|,
\]
is minimized among all branches with the same $z_{\rm inf}$ in $S'$.
\item 
The redshift, $z_{\rm merge}'$, at which $B'$ merges into another subhalo in $S'$,
is compared with the redshift, $z_{\rm merge}$, at which $B$ merges into another 
subhalo in $S$. If and only if $z_{\rm merge}' < z_{\rm merge}$, 
the lifetime of $B$ is extended to $z_{\rm merge}'$.
\item 
If $B$ is extended, a list of new subhalo is created accordingly and 
attached to the tree.
\eenum

Once the central-stage and satellite-stage completion steps are taken,  
links between subhalos in merger trees of $S$, such as the progenitor and 
descendant relationships, as well as group memberships, are updated to 
reflect the extension.

\subsubsection{Phase-space Assignment} 
\label{sssec:phase-space-assignment}

In the phase-space assignment step, we assign positions 
and velocities to all extended satellite subhalos in $S$. 
The phase-space properties of a satellite subhalo are expected to be correlated 
with other properties. For example, a satellite subhalo of earlier 
infall is expected to have higher probability to appear in the 
inner region of its host halo, while a subhalo of recent infall 
is expected to reside in the outskirt. 
Other studies have also shown that some properties at the infall time of 
a satellite subhalo, such as the orbital angular momentum and its mass 
ratio with the central subhalo, are the main factors that affect its orbital 
dynamics \citep[e.g.,][]{boylan-kolchinDynamicalFrictionGalaxy2008}.
Because of these correlations, it is possible to design an algorithm 
that not only assigns positions and velocities randomly to satellite 
subhalos, but can also recover the distribution of the satellite population, 
$p({\bf x}_{\rm sat})$, with respect to a set of variables, 
${\bf x}_{\rm sat}$, such as position, velocity, and other properties.

In general, modeling the full probability density function (PDF)
of ${\bf x}_{\rm sat}$ is challenging due to its high dimensionality. 
To simplify the problem, we split ${\bf x}_{\rm sat}$ into two subsets 
of variables, ${\bf x}_{\rm sat, complete}$, which can be completely resolved in 
$S$, and ${\bf x}_{\rm sat, incomplete}$, which is missed for some 
subhalos in $S$ and needs to be assigned. 
We use the following constraints in the splitting:
\benum
\item 
The incomplete set ${\bf x}_{\rm sat, incomplete}$ must include position 
and velocity, or some transformations of them, because they are missed 
for subhalos in the extension and are the target properties of this step.
\item 
The spatial distribution of satellite subhalos must be compliant with 
the constraints imposed by their host halos. For example, 
theoretical and numerical studies both show that halos tend to 
be ellipsoidal rather than spherical 
\citep[e.g.][]{shethEllipsoidalCollapseImproved2001,
maccioConcentrationSpinShape2007,
chenRelatingStructureDark2020}, and so satellite subhalos
are also expected to have non-spherical distribution if they trace the 
density field in their host halos. 
This anisotropy are clearly seen in the distribution of 
simlulated satellites shown in Fig.~\ref{fig:extension-spatial-points}. 
Thus, to better recover subhalo distributions in individual 
host halos, the extension algorithm should make use of shape information 
of halos, namely it should be ``shape-preserving''.
\item 
Because many satellite subhalos are resolved in $S$, 
as can be seen from Fig.~\ref{fig:extension-hmf}, 
the algorithm is required to retain their ${\bf x}_{\rm sat, incomplete}$ 
given by $S$ as long as this does not break any consistency with the 
distribution of ${\bf x}_{\rm sat}$ obtained from $S'$. 
This requirement implies that the ``retained'' subhalos are not 
only a statistically valid population, but also compliant to 
$S$ on a per-subhalo basis. The use of properties given by 
$S$ in the extension algorithm is referred to as ``self-consistency''.
\eenum

Once the split is made for ${\bf x}_{\rm sat}$, we can use 
the product rule of probability to decompose the full PDF into two terms:
\begin{equation}
    p({\bf x}_{\rm sat}) = p({\bf x}_{\rm sat, complete}) 
        p({\bf x}_{\rm sat, incomplete} | {\bf x}_{\rm sat, complete}),
\end{equation}
where the first and second factors on the right hand side are 
the conditioning and conditioned terms, respectively. The first term can be 
estimated reliably from $S$ as a result of the definition of 
${\bf x}_{\rm sat, complete}$. The conditioned term, on the other hand, 
is unknown from $S$, and has to be derived elsewhere, for example, from $S'$.
This decomposition strategy has been widely adopted in theoretical 
modeling of halos and galaxies. For example, HOD models mainly target at the 
number of member galaxies of a host halo conditioned on the halo mass. 
The conditional luminosity functions (CLFs), conditional galaxy stellar mass 
functions, and conditional HI mass functions (CHIMFs) 
extend this and model respectively 
the distributions of galaxy luminosity, stellar mass, and HI gas mass, 
conditioned on halo mass \citep{yangConstrainingGalaxyFormation2003,
zandivarezLuminosityFunctionGalaxies2006,
yangGalaxyGroupsSDSS2008,
robothamVariationGalaxyLuminosity2010,
zandivarezLuminosityFunctionGalaxies2011,
lanGalaxyLuminosityFunction2016,mengGalaxyPopulationsGroups2022,
liConditionalHIMass2022}. This idea is also 
used by \citet{chenELUCIDVICosmic2019} to fix the cosmic variance at the 
low-stellar-mass end of the galaxy stellar mass function. 
The CCMD model of \citet{xuConditionalColourMagnitude2018} 
further extends the conditional distribution by including both 
magnitude and color as targets. The difference in our task is that the 
conditioning variable ${\bf x}_{\rm sat, complete}$ is mutivariant, and hence, the 
computation and application of $p({\bf x}_{\rm sat, incomplete} | {\bf x}_{\rm sat, complete})$
require {\revstyle partitions} in a high-dimensional feature space. To tackle this, 
we design the following substeps to numerically 
learn the conditioned distribution from $S'$ and assign 
phase-space properties to satellites in $S$ according to the results learned.
\benum
\item 
We compute ${\bf x}_{\rm sat, complete}$ for all satellite subhalos
in both $S$ and $S'$, and we compute ${\bf x}_{\rm sat, incomplete}$
for all satellite subhalos in $S'$ and all simulated satellite subhalos 
in $S$. In addition, for any subhalo in $S$, a binary variable, 
$I_{\rm missed}$, is defined to indicate whether or not it is missed by 
the simulation and thus created in the step of satellite-stage completion.
\item 
We train a CART tree classifier \citep{breimanClassificationRegressionTrees1984} 
that maps ${\bf x}_{\rm sat, complete}$ to $I_{\rm missed}$. Here, the objective function 
is the misclassification rate and the training sample consists of 
satellite subhalos from $S$. So trained, the feature space of 
${\bf x}_{\rm sat, complete}$ is partitioned 
into a set of subregions $\{C_i\}_{i=1}^{N_{\rm cell}}$ by 
the CART tree, with time-integrated effects of environment
naturally taken into account. Internally, the CART tree represents each 
subregion $C_i$ by one of its leaf nodes, and makes prediction for 
a test data point according to the subregion the point is located in.
In what follows, we refer to each subregion as a ``cell'' and we use $N_{\rm cell}$ 
to denote the total number of cells. To alleviate effects of 
overfitting due to cosmic variances, we control the fineness of the partition 
in the training process by limiting the number of 
subhalos in each cell to be no less than a minimal value, 
$N_{\rm min, cell\ partition}$, and the total number of cells to be no 
larger than a maximal value, $N_{\rm cell, max}$.
\item 
Satellite subhalos in $S$ and $S'$ are assigned to cells according 
to their ${\bf x}_{\rm sat, complete}$. In each cell $C_i$, subhalos from 
$S$ and $S'$ are collectively denoted as $H_i$ and $H_i'$, respectively:
\begin{align}
    H_i  &= \{h \in S | {\bf x}_{\rm sat, complete}(h) \in C_i \}, \\
    H_i' &= \{h \in S' | {\bf x}_{\rm sat, complete}(h) \in C_i \},
\end{align}
where $h$ denotes a satellite subhalo.
\item 
The location of $H_i$ (or $H_i'$) in the feature space is defined by 
averaging ${\bf x}_{\rm sat, complete}$ among all subhalos in it:
\begin{align}
    {\bf x}_{\rm sat, complete}({H_i}) &= \frac{1}{N_{H_i}} \sum_{h \in H_i} {\bf x}_{\rm sat, complete}(h), \\ 
    {\bf x}_{\rm sat, complete}({H_i'}) &= \frac{1}{N_{H_i'}} \sum_{h \in H_i'} {\bf x}_{\rm sat, complete}(h),
\end{align}
where $N_{H_i}$ and $N_{H_i'}$ are the numbers of subhalos in $H_i$ and $H_i'$, respectively. 
\item 
We perform a ``cell-matching'' that identifies, for each 
$H_i$ ($1 \leqslant i \leqslant N_{\rm cell}$), a closest
neighbor from $H_j'$ ($1 \leqslant j \leqslant N_{\rm cell}$).
Specifically, for each cell $C_i$, $H_i$ is matched with $H_i'$
if $N_{H_i'}$ is larger than a predefined threshold, $N_{\rm min, cell\ match}$.
Otherwise, $H_i'$ is considered too small to provide a robust  
estimate of the PDF of ${\bf x}_{\rm sat, incomplete}$ in that cell, 
and we use the NNM to search for a $H_j'$ in another cell 
$C_j$ to identify the $H_j'$ that is closest to $H_i$
according the $L_2$ distance,  
\begin{equation}
    d_{\rm cell}(H_i, H_j') = \| {\bf x}_{\rm sat, complete}({H_i}) - 
        {\bf x}_{\rm sat, complete}({H_j'}) \|,
\end{equation}
and has $N_{H_j'} \geqslant  N_{\rm min, cell\ match}$.
With such cell-matching, each cell $C_i$ is attached with 
a sufficiently large sample of subhalos from $S'$, so that 
we can estimate robustly the PDF,
$p({\bf x}_{\rm sat, incomplete} | C_i)$, conditioned in this cell. 
This PDF will be used as an approximation to the exact PDF
$p({\bf x}_{\rm sat, incomplete} | {\bf x}_{\rm sat, complete})$
for any ${\bf x}_{\rm sat, complete} \in C_i$.
\item 
For each cell $C_i$, we perform a ``conditional abundance matching''
to assign a ${\bf x}_{\rm sat, incomplete}$ to each subhalo in $H_i$, 
using the properties of its closest match in $H_j'$. 
The quantities used to match and the order of matching 
depend on the details of $S$ and $S'$ and on the exact set of 
properties to be borrowed from $S'$ and assigned to $S$. 
Independent of the detail, 
the general constraints are that the conditional distribution, 
$p({\bf x}_{\rm sat, incomplete} | C_i)$, must be recovered in $H_i$ 
after the assignment, and that the assignment is shape-preserving 
and self-consistent, as stated at the beginning of this step.
\eenum

With all these steps, an extended version of subhalo merger trees 
is obtained for $S$.

\subsection{Application to ELUCID and TNGDark}
\label{ssec:specific-elucid-and-tngdark}

\begin{figure*} \centering
    \includegraphics[width=0.85\textwidth]{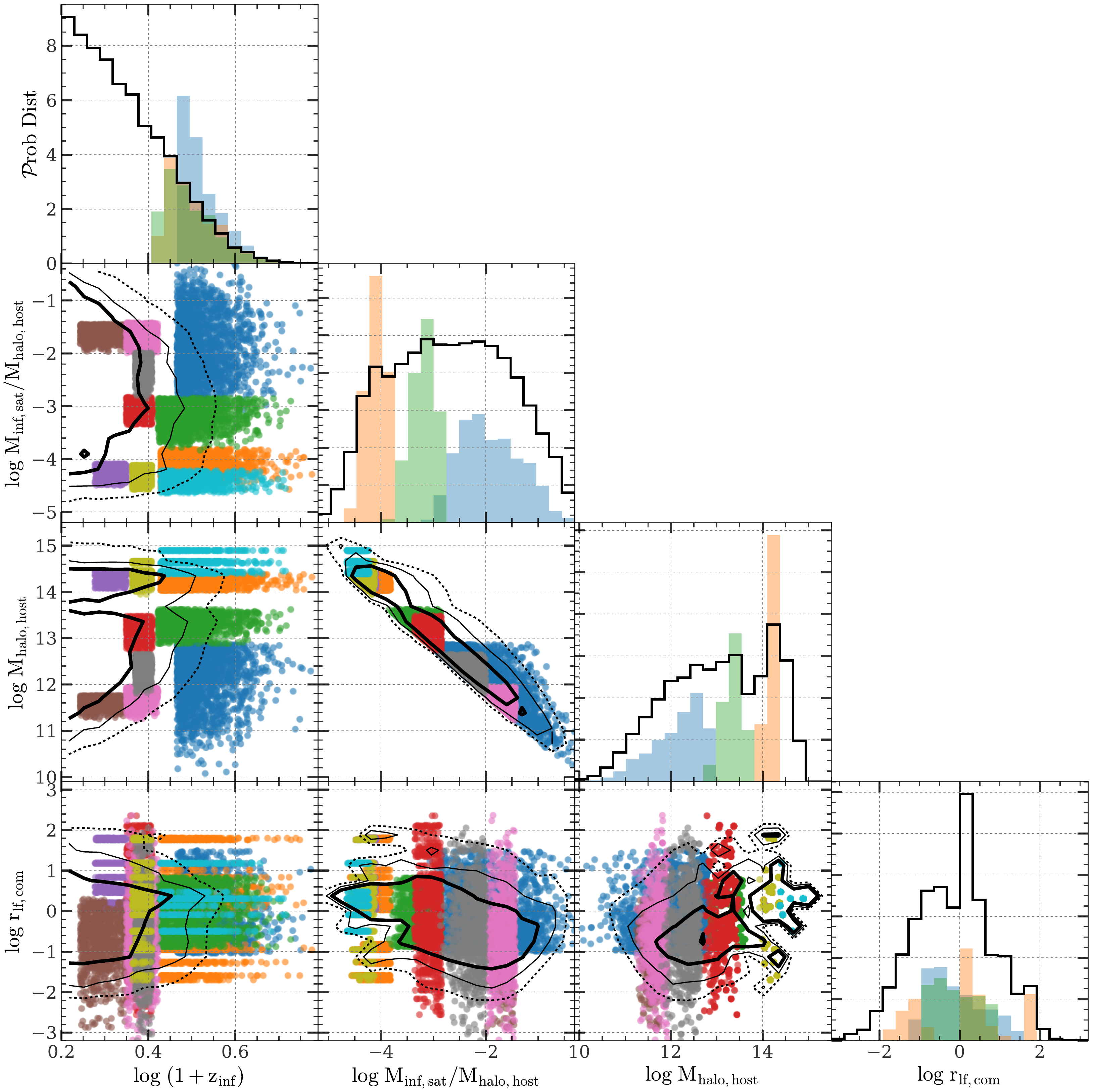}
    \caption{
        Marginal distributions of $z=0$ ELUCID satellite subhalos
        in the projected spaces of properties that are used as the conditioning variables
        in the phase-space assignment step (see Table~\ref{tab:specific-choices} and 
        \S\ref{ssec:specific-elucid-and-tngdark} for details). 
        Satellite 
        subhalos that are resolved by ELUCID and created in the satellite-stage 
        completion step are both included.
        Each diagonal panel shows the 1-D
        distribution of a property. Each off-diagonal 
        panel shows the distribution of a pair of properties.
        In each diagonal panel, the black histogram shows the distribution 
        of all satellite subhalos while a colored histogram show the distribution
        of subhalos in a cell found by the CART tree. Only the biggest 
        three cells are shown. 
        The histograms are arbitrarily normalized for clarity.
        In each off-diagonal panel, the black thick solid, thin solid and dotted lines 
        are contours enclosing $50\%$, $75\%$ and $90\%$ of all satellite subhalos, 
        respectively. 
        Dots with the same color represent subhalos belonging to the same cell.
        The biggest 10 cells are shown.
    }
    \label{fig:extension-cells}
\end{figure*}

In this application, we extend subhalo merger trees in $S = {\rm ELUCID}$.
Here we first specify choices of reference simulation, computation strategies, 
subhalo quantities and algorithm parameters for this specific application.  

As shown by \citet{vandenboschDarkMatterSubstructure2018} with a suite of 
idealized simulations, satellites are easily affected by numerical defects even
with large number of bound particles. They found that reliably resolving the tidal 
evolution of a satellite for a Hubble time on a circular orbit at $20\%$ ($10\%$) of 
the virial radius of the host halo requires $10^5$ ($10^6$) particles. 
This is too demanding for any state-of-the-art cosmological simulation.
For the problem tackled here, because we only require the satellite disruption 
time and phase-space properties be statistically correct in the reference
simulation $S'$, a more relaxed condition may be sufficient.
As shown by \citet{hanUnifiedModelSpatial2016} with 
a suite of realistic zoom-in simulations, the number density profile for 
resolved satellites increases with numerical resolution and becomes convergent 
when $N_{\rm acc}$, the minimal particle number of satellite at accretion, 
is larger than $\sim 10^3$. The same conclusion was reached by 
\citet{guoNumericalResolutionLimits2013} using the TPCFs of galaxies 
predicted by applying the subhalo abundance matching technique 
to a pair of simulations with different numerical resolutions. 
If we adopt $N_{\rm acc} = 10^3$ for the least massive satellite
in ELUCID ($M_{\rm inf} \sim 10^{10} \msun$), the reference 
simulation $S'$ is required to have a particle mass less 
than $10^7 \msun$. 
Based on these, our choice of $S'={\rm TNGDark}$ 
as the reference simulation is appropriate for extending ELUCID.
{\revstyle
In Appendix~\ref{app:ssec:convergence}, we present a convergence analysis for 
the volume of the reference simulation. Our findings indicate that the size of 
the TNGDark volume is sufficiently large to encompass a representative population 
of (sub)halos needed for the extension algorithm.
}

To tackle the large data volume of ELUCID, we split the simulation box 
of $(500 \mpc)^3$ volume into $5\times 5 \times 5$ equal-sized, non-overlapping 
subboxes, each with volume of $(100 \mpc)^3$. We run the extension algorithm for 
each subbox independently, and combine the resulted merger trees from all subboxes 
into a final data product. With such implementation, the required memory and 
computation costs of each subbox are reasonable for a single node of a 
modern computer, and the computation in different subboxes can be made  
parallel with a cluster of nodes.

For the central-stage completion step, we define 
${\bf x}_{\rm brh, cent}$, the set of properties to be used
in matching branches between $S$ and $S'$, as
\begin{equation}
\label{eq:specific-choice-on-x-brh-cent}
    {\bf x}_{\rm brh, cent} = \,[\log M_{\rm halo,inf},\, \log(1+ z_{1/2}) \,]
\end{equation}
for all branches with $M_{\rm halo,inf} \geqslant M_{\rm match, cent}$, and
\begin{equation}
    {\bf x}_{\rm brh, cent} = \log M_{\rm halo,inf}
\end{equation}
for all branches with $M_{\rm halo,inf} < M_{\rm match, cent}$.
The parameter $M_{\rm match, cent}$ has to be chosen so that branches with 
$M_{\rm halo,inf} \geqslant M_{\rm match, cent}$ have reliable values of 
$z_{\rm 1/2}$ in $S$. For $S={\rm ELUCID}$, we have made tests and found that 
$M_{\rm match, cent} = 2 \times 10^{10} \msun$, the mass of about 60 N-body 
particles, is an appropriate choice. Similarly, we set 
$M_{\rm lim, cent}=10^{10} \msun$, which defines the joint redshift 
$z_{\rm joint}$ of each branch in $S$ in extending the central part of the MAH. 
Because $M_{\rm halo,inf}$ and $z_{1/2}$ describe the overall amplitude 
and detailed shape of the MAH, respectively, our choice ensures that 
${\bf x}_{\rm brh, cent}$ is tightly correlated with the MAH. 
Our tests show that this produces a smoother transition at the joint 
redshift $z_{\rm joint}$ for individual subhalos than the simple method 
used by \citet{chenELUCIDVICosmic2019}. Using a demarcation of infall mass 
at $M_{\rm match, cent}$, we split branches in each of $S$ and $S'$ 
into two sub-samples. For the higher-mass and lower-mass sub-samples of $S$, we 
use the higher-mass and lower-mass sub-samples of $S'$, respectively, 
to accomplish the central-stage completion. To suppress distribution shift 
produced by potential discrepancy between the two simulations, we standardize 
${\bf x}_{\rm brh, cent}$ and ${\bf x}_{\rm brh, cent}'$
so that they have 
zero mean and unit standard deviation along all dimensions before applying 
the NNM.

To accomplish the satellite-stage completion, we need to specify the 
set of branch properties, ${\bf x}_{\rm brh, sat}$, to be used to match branches 
between $S$ and $S'$. Here, we choose
\begin{equation}
    {\bf x}_{\rm brh, sat} = (\log M_{\rm halo, inf},\,\log M_{\rm halo, cent, inf}, 
        \, \log\,j_{\rm inf}),
\end{equation}
where $M_{\rm halo, inf}$ and $M_{\rm halo, cent, inf}$ are the infall mass of 
the satellite subhalo and the mass of the host halo it is falling into, 
respectively, and $j_{\rm inf}$ is the orbital angular momentum. This choice 
is motivated by the fact that these properties dominate the orbital dynamics 
of a satellite subhalo \citep[see, e.g.,][]{boylan-kolchinDynamicalFrictionGalaxy2008},
and that these properties are numerically stable \citep[see, e.g., Figure A3 in][]{chenMAHGICModelAdapter2021}.
Similar choices have been adopted in some previous empirical models of galaxy 
formation, such as those developed by 
\citet{luEmpiricalModelStar2014,luStarFormationStellar2015}.
As in the central-stage completion, standardization of 
${\bf x}_{\rm brh, sat}$ is made before applying the NNM 
to suppress distribution shift caused by potential discrepancy 
between the two simulations.

In the step of assigning phase-space coordinates to satellite subhalos,
diversity of dark matter halo properties such as mass, size, shape and
orientation requires a large set of halo properties to be included 
in ${\bf x}_{\rm sat, complete}$ in order to reliably model the conditional 
PDF, $p({\bf x}_{\rm sat, incomplete} | {\bf x}_{\rm sat, complete})$. 
Such a model is in general very complicated. Here we simplify 
the problem by reducing the number of variables. 
To this end, we transform the phase-space properties of a satellite subhalo
using the properties of its host halo, so that they are scaled by  
the ``local frame'' defined by the host. By so doing, the host properties 
are eliminated from the conditioning variable ${\bf x}_{\rm sat, complete}$, 
and the conditioned variable ${\bf x}_{\rm sat, incomplete}$ 
becomes dimensionless. 
This is, effectively, a stacking method that first scales the properties 
in different systems and then combines the scaled quantities to enhance 
the signal. This method has been used frequently in literature to 
extract features from weak signals, such as images or spectra with low 
signal-to-noise ratios. 

For each host halo, we first compute its inertial tensor $\mathcal{I}$ using
\begin{equation}
\label{eq:inertial-tensor}
    \mathcal{I} = 
        \frac{1}{2} m_{\rm p}
        \sum_i \Delta {\bf r}_{\rm p, i}\,  \Delta {\bf r}_{\rm p, i} ^T, \\
\end{equation}
where the summation is over all the $N_{\rm p}$ dark matter particles belonging to the halo, 
$\Delta {\bf r}_{\rm p, i} = {\bf r}_{\rm p, i} - {\bf r}_{\rm com}$ is the 
position vector of the $i$-th particle relative to the center of mass (COM), 
${\bf r}_{\rm com} = \frac{1}{N_{\rm p}} \sum_i {\bf r}_{\rm p, i} $, and 
$m_{\rm p}$ is the mass of each particle.
Then, we compute the eigenvalues, $\lambda_i$, and eigenvectors, ${\bf e}_i$, 
of the inertial tensor. 
We describe the shape of the halo by the principal axes, $a_i\,(i=1,2,3)$, 
of its inertial ellipsoid:
\begin{equation}
\label{eq:eigen-axes}
    a_i = \sqrt{\lambda_i}.
\end{equation}
The eigenvectors and the principal axes define the local frame of the halo, 
to which we tranform the position, $\bf r$, and velocity, $\bf v$, 
of each member subhalo using
\begin{align}
    {\bf r}_{\rm lf} &= R_{\rm halo, host}^{-1} \mathcal{S E} ({\bf r} - {\bf r}_{\rm com}), \nonumber \\
    {\bf v}_{\rm lf} &= V_{\rm halo, host}^{-1} \mathcal{E} ({\bf v}-{\bf v}_{\rm com}).  
\label{eq:local-frame-transformation}
\end{align}
Here $R_{\rm halo, host}$ and $V_{\rm halo, host}$ are the virial radius and virial 
velocity of the host halo, respectively; 
${\bf v}_{\rm com} = \frac{1}{N_{\rm p}} \sum_i {\bf v}_{\rm p, i} $ 
is the velocity of the COM obtained by averaging the velocities of  
all particles in the halo; $\mathcal{E} = ({\bf e}_1, {\bf e}_2, {\bf e}_3)^T$ 
is the rotational matrix; $\mathcal{S} = {\rm diag}(s_1, s_2, s_3)$ is the 
stretching matrix along the three principal axes, with the stretching 
factor $s_i$ along the $i$-th principal axis defined as
\begin{equation}
\label{eq:stretch-factor}
    s_i = {(a_1 a_2 a_3)^{1/3}\over a_i}\,.
\end{equation}

To describe the radial and angular distribution of satellite subhalos in the 
local frame defined by the host halo, we define, for a subhalo located  
at ${\bf r}_{\rm lf}$ with velocity ${\bf v}_{\rm lf}$, its halo-centric 
distance $r_{\rm lf}$ and position angle $\theta_{r, {\rm lf}}$ as
\begin{align}
    r_{\rm lf} &= \| \Delta {\bf r}_{\rm lf} \|, \nonumber \\
    \cos{ \theta_{r, {\rm lf}} } &= \Delta {\bf r}_{\rm lf}
        \cdot \frac{
            \Delta {\bf r}_{\rm lf, com}
        }{
            \| \Delta {\bf r}_{\rm lf, com} \|
        }\,.
\label{eq:spherical-coord-of-x}
\end{align}
Here,  
$\Delta {\bf r}_{\rm lf} \equiv {\bf r}_{\rm lf} - {\bf r}_{\rm lf, cent}$, and
$\Delta {\bf r}_{\rm lf, com} \equiv {\bf r}_{\rm lf, com} - {\bf r}_{\rm lf, cent}$,
with ${\bf r}_{\rm lf, cent}$ and ${\bf r}_{\rm lf, com}$ being the 
local-frame positions of the central subhalo and the COM of the host 
halo, respectively. So defined, $r_{\rm lf}$ and $\theta_{r, {\rm lf}}$ are, 
respectively, the radial distance and polar angle in the spherical 
coordinate system with the polar axis parallel to $\Delta{\bf r}_{\rm lf, com}$.

Similarly, we define the halo-centric speed
$v_{\rm lf}$ and velocity polar 
angle $\theta_{v, {\rm lf}}$ as 
\begin{align}
    v_{\rm lf} &= \| \Delta {\bf v}_{\rm lf}  \|, \nonumber \\
    \cos{ \theta_{v, {\rm lf}} } &= 
        \Delta {\bf v}_{\rm lf}
        \cdot \frac{
            \Delta {\bf v}_{\rm lf, com}
        }{
            \| \Delta {\bf v}_{\rm lf, com} \|
        }\,,
\label{eq:spherical-coord-of-v}
\end{align}
where $\Delta {\bf v}_{\rm lf} \equiv {\bf v}_{\rm lf} - {\bf v}_{\rm lf, cent}$, and 
$\Delta {\bf v}_{\rm lf, com} \equiv {\bf v}_{\rm lf, com} - {\bf v}_{\rm lf, cent}$.
${\bf v}_{\rm lf, cent}$ and ${\bf v}_{\rm lf, com}$
are the local-frame velocities of the central subhalo 
and of the COM of the host halo, respectively.
Note that
both ${\bf r}_{\rm lf, com}$ and ${\bf v}_{\rm lf, com}$ are zero
by their definitions.

With phase-space properties defined in the local frame, we choose the 
properties in the conditional PDF of the phase-space assignment step as
\begin{align}
\label{eq:conditioning-and-conditioned-vars}
        {\bf x}_{\rm sat, complete} =& \,[\
            \log (1+z_{\rm inf}), \ \log \frac{M_{\rm inf, sat}}{M_{\rm halo, host}}\,, \nonumber \\
            &\ \ \ \ \ \ \  \ \ \ \ \ \ \ \log M_{\rm halo, host}, \ r_{\rm lf, com}
        \ \,], \nonumber \\
        {\bf x}_{\rm sat, incomplete} = & \, ({\bf r}_{\rm lf},\,{\bf v}_{\rm lf})\,.
\end{align}
Here, $z_{\rm inf}$ and $M_{\rm inf, sat}$ are the infall redshift and infall 
mass of the satellite subhalo, respectively, and 
$M_{\rm halo, host}$ is the current mass of the host halo. 
The separation, $r_{\rm lf, com} \equiv \| \Delta {\bf r}_{\rm lf, com}  \|$, is 
a quantity that measures the relaxation state of the host subhalo 
\citep[see, e.g.,][]{maccioConcentrationSpinShape2007,
ludlowDynamicalStateMassConcentration2012, chenRelatingStructureDark2020},
and is included here to control un-relaxed systems that are expected to  
be more asymmetric in their mass distribution (see \S\ref{ssec:individual-halos} 
and Fig.~\ref{fig:extension-spatial-points} for some examples). 
By using ${\bf r}_{\rm lf}$ and ${\bf v}_{\rm lf}$ as target variables, 
the shape information of the host halo is automatically included.
In some cases, for example, when simulating an extremely large volume or 
simulating a large ensemble of volumes, storing the full catalog of dark matter 
particles into disk is infeasible. Then, we can simply 
remove the shape informaton and degrade the local frame (Eq.~\ref{eq:local-frame-transformation})
to a spherically symmetric coordinate system. Our tests show that, with this
simplification, the shape-preserving feature is lost, but spherically averaged 
summary statistics, such as the number density profiles for satellites and the 
TPCFs for subhalos, are still precisely corrected by the extension algorithm. 

When using the CART tree to split the feature space of
${\bf x}_{\rm sat, complete}$ in cells, we need to specify a stopping 
criterion for the recursive space partitioning. Throughout this paper, we set 
$N_{\rm cell, max}=768$ and $N_{\rm min, cell\ partition} = 32$, which 
gives the upper bound of the number of cells and the lower bound of 
the number of satellite subhalos in each cell, respectively. 
We have made tests by allowing a relatively large $N_{\rm cell, max}$, 
and found that the partition of feature space is sufficiently fine to 
reproduce the joint distribution of satellite properties we are interested in.
By limiting the minimal cell size, the uncertainties caused by the cosmic variance 
can be controlled effectively, thus making the extension more stable.
With a similar consideration,  we set $N_{\rm min, cell\ match}=32$,
which gives the lower bound of the number of satellites from $S'$ 
in the matched  cell. Note that these values are specific to the 
simulations used here, and should be tested when applying the method 
to other datasets.

Fig.~\ref{fig:extension-cells} shows the distribution of satellite subhalos
from the first subbox of ELUCID in projected spaces of the conditioning 
variable ${\bf x}_{\rm sat, complete}$.
Subhalos in several largest cells are plotted using colored points.   
In all 2-D panels, cells are regular rectangles because of of the bi-partition 
nature of the CART tree classifier.
The 1-D distribution of the host halo mass, $\log M_{\rm halo, host}$, shows 
a concentration at $10^{14} \msun$, indicating a significant cosmic variance 
in the ELUCID subbox used here. 
Several largest cells, such as those colored with cyan, orange, yellow and 
purple, are located in the this concentration
This indicates that the classifier captures this special population 
of satellites in massive halos where environmental effects 
are strong, and allocates individual cells to them. 
Some horizontal strips are clearly seen in the 2-D plots, 
because massive halos are rare and all satellites in one such halo 
share the same $M_{\rm halo, host}$ and $r_{\rm lf, com}$.
Cells are well separated in the 2-D panels along the axes of 
$\log (1+z_{\rm inf})$, 
$\log \frac{M_{\rm inf, sat}}{M_{\rm halo, host}}$ and
$\log M_{\rm halo, host}$, indicating the importance of 
these variables in predicting numerical defects 
indicated by $I_{\rm missed}$ 
\citep[see, e.g.,][]{boschDisruptionDarkMatter2018,greenTidalEvolutionDark2021}.
This is expected, because environmental processes, no matter physical or 
numerical, have time-integrated effects that depend on the potential of 
the satellite itself, the density and tidal strength of the 
host halo, and the time duration since the infall.
In contrast, significant overlaps of cells are seen along the axis of 
$r_{\rm lf, com}$, indicating that 
incomplete relaxation of host halos has a more subtle effect on satellite dynamics.

Finally, we specify our choice to rank order features used in   
the conditional abundance matching.
We choose the halo-centric distance, $r_{\rm lf}$, as the 
target variable to match, because radial distributions of 
satellite subhalos in their host halos are the main targets we want to 
reproduce, and because the polar angle, $\theta_{\rm lf}$, 
is not significantly correlated with $r_{\rm lf}$, as seen 
from Fig.~\ref{fig:extension-2d-hist} that will be described in 
detail later. With this choice, the matching algorithm
proceeds for each cell $C_i$ in the following substeps:
\benum
\item 
We collect the set of $r_{\rm lf}$ values from all ELUCID-simulated 
satellite subhalos that fall into the cell, and denote it as $R$:
\begin{equation}
    R = \{ r_{\rm lf}(h)\, |\, h \in H_i\ {\rm and} \ I_{\rm missed} = 0 \}.
\end{equation}
Similarly, the set of $r_{\rm lf}$ values in the matched cell from TNGDark 
is denoted as $R'$:
\begin{equation}
    R' = \{ r_{\rm lf}(h)\, |\, h \in H_j' \}.
\end{equation}
\item 
We re-sample $R'$ so that the size of the re-sampled set is equal to the 
size of $H_i$. If the original size of $R'$ is less than required, 
the resampling has replacement; otherwise it does not.
\item 
For each simulated ELUCID satellite with a halo-centric distance 
${r_{\rm lf}} \in R$, we match it with a TNGDark satellite  
that has a halo-centric distance ${r_{\rm lf}}' \in R'$, requiring that
\begin{equation}
\label{eq:cond-matching-radial-criterion}
    {\rm \Delta}\log r_{\rm lf} \equiv |\log\,r_{\rm lf} 
        - \log\,r_{\rm lf}'| \leqslant {\rm \Delta}\log r_{\rm lf, max}.
\end{equation}
where $\Delta \log r_{\rm lf, max}$ limits the matching range and is set to be
$0.1$. The matching starts from the most massive satellite,
as measured by $M_{\rm inf, sat}$, in ELUCID, to the least massive one.
If multiple satellites are found in TNGDark for an ELUCID satellite, 
the one with the smallest ${\rm \Delta}\log r_{\rm lf}$ is selected. Once a match is found, 
the matched satellite in TNGDark is removed from $R'$; otherwise, no  
match is made, and we continue with the next ELUCID satellite.
\item 
For each ELUCID satellite that is matched with TNGDark satellite, 
we set its phase-space properties, $({\bf r}_{\rm lf}, {\bf v}_{\rm lf})$, 
to the simulated values in ELUCID. We refer to these satellites as 
``ELUCID satellites'', and their mass function is shown by the blue solid line in 
Fig.~\ref{fig:extension-hmf}. For comparison, the blue dashed line
in that figure accounts for all satellites resolved in ELUCID without regard 
to the matching.
\item 
For the remaining ELUCID satellites, 
either created in the satellite-stage completion step or 
unmatched to any TNGDark satellite in the previous substep,
we randomly match them, one-to-one, with TNGDark satellites 
that have $r_{\rm lf}'$ values in $R'$. 
We use $(r_{\rm lf}, \theta_{r, {\rm lf}})$ and  
$(v_{\rm lf}, \theta_{v, {\rm lf}})$
from the matched TNGDark subhalo, together with randomly generated 
azimuthal angles $\phi_{r, {\rm lf}}$ and $\phi_{v, {\rm lf}}$ 
(respectively for the position and velocity) to obtain the local-frame 
coordinates $({\bf r}_{\rm lf}, {\bf v}_{\rm lf})$ for each of the 
remaining ELUCID satellites. These ELUCID satellites are referred 
to as the population of extension, and their mass function 
is shown by the red solid line (labeled ``Extension'')  
in Fig.~\ref{fig:extension-hmf}. For comparison, the red dashed line 
is the result for satellites that are created in the step 
of satellite-stage completion.
\eenum 

Once an ELUCID satellite subhalo is assigned values of 
$({\bf r}_{\rm lf}, {\bf v}_{\rm lf})$ either directed
by the target simulation or by the extension algorithm, 
its physical coordinates in phase-space can be obtained 
by inverting the transformations represented by 
Eq.~\ref{eq:local-frame-transformation}.

\begin{figure*} \centering
    \includegraphics[width=\textwidth]{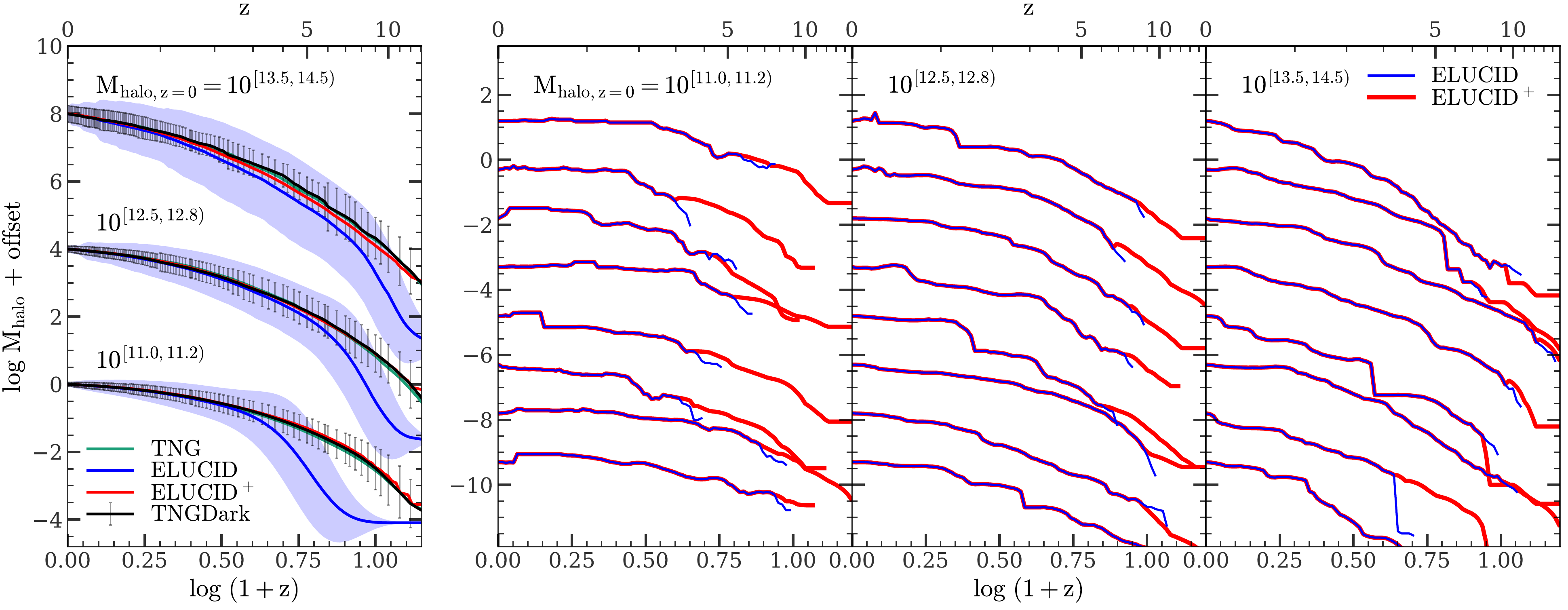}
    \caption{
        Mass assembly histories of central subhalos at $z=0$ in different
        simulations. 
        Curves are shown with different offsets for clarity.
        The leftmost panel shows the average histories of subhalos in three  
        bins of $M_{\rm halo,z=0} / (\msun)$, indicated above 
        each bunch of curves. 
        Green, blue, red and black 
        lines are mean values from TNG, ELUCID, $\rm ELUCID^+$ and TNGDark, 
        respectively, with black errorbars and blue shaded areas indicating 
        the corresponding standard deviations among branches.
        The right three panels show the assembly histories of individual 
        subhalos randomly selected in three bins of $M_{\rm halo, z=0}/(\msun)$, 
        respectively, indicated at the top of the panels. For each subhalo, 
        blue line shows its assembly history before the central-stage 
        completion, which is truncated near the resolution limit of ELUCID. 
        The red line shows the result after extension, which smoothly continues
        to the mass limit defined by the reference simulation, TNGDark.
    }
    \label{fig:extension-central-history}
\end{figure*}

\section{Testing the Performance of the Extension Algorithm}
\label{sec:results}

The extension algorithm developed above produces subhalo merger trees 
that are more complete in MAH for both the central and satellite subhalo 
populations. Because the extension is shape preserving and self-consistent, 
the trees also retain important information contained in the original, target 
simulation. In this section, we present various testing results to demonstrate 
the reliability and accuracy of the extension algorithm.

\subsection{Mass Assembly Histories of Central Subhalos}
\label{ssec:test-mah}

The central-stage completion step (\S\ref{sssec:central-stage-completion}) 
of our algorithm completes the assembly
histories of central subhalos at high redshift when their masses 
are too small to be resolved in the target simulation. 
Fig.~\ref{fig:extension-central-history} compares  
the MAHs obtained from the target simulation 
(ELUCID), the extended version of it ($\rm ELUCID^+$), 
the reference simulation (TNGDark), and the hydro counterpart 
of the reference simulation (TNG). 
The leftmost panel shows the average MAHs of branches with 
$z_{\rm inf} = 0$ in three bins of halo masses at $z=0$. 
We can understand the result using a series of 
pair-wise comparisons. First, the MAHs of TNG are almost indistinguishable 
from those of TNGDark up to $z \sim 10$. This indicates that the overall 
halo properties, such as the virial mass and the virial radius, 
are stable against baryonic effects. This stability forms the 
basis for empirical models built on DMO simulations.
Second, significant discrepancy can be seen between TNGDark and ELUCID
at $z \gtrsim 2$. Above this redshift, the average MAHs of ELUCID
are gradually dominated by unresolved, low-mass subhalos whose MAHs 
are padded  artificially. Thus,
an empirical model based on the MAHs of such incomplete histories
will miss star formation in low-mass halos at high redshift.
Finally, with the central-stage completion, the average MAHs 
of $\rm ELUCID^+$ become consistent with the high resolution 
simulation TNGDark over the entire redshift range shown. 
This indicates that our NNM-based extension produces unbiased
MAH in the full redshift range up to $z \sim 10$.
The standard deviations of the MAHs in ELUCID, shown by 
the blue shaded areas, are larger than those in TNGDark, 
even at low $z$. This is a result of the variation in the first 
resolvable redshift, $z_{\rm first}$, of ELUCID branches.

The right block of three panels in Fig.~\ref{fig:extension-central-history}
shows the MAHs of individual branches randomly selected in three different 
mass bins at $z=0$. The MAHs resolved by ELUCID are all truncated 
at some redshifts when their masses are below the resolution limit of ELUCID. 
In contrast, the MAHs after the extension, labeled as $\rm ELUCID^+$, 
start to deviate from those of ELUCID at some joint redshifts 
$z_{\rm joint}$, but extend smoothly to higher redshift, eventually  
being truncated as their mass goes below an effective 
mass limit defined by TNGDark.
We note that the smoothness around $z_{\rm joint}$ is a 
combined outcome of the discontinuity-removal applied in the substep (iii)
of the central-stage completion, and the specific choice of 
${\bf x}_{\rm brh,cent}$ made for ELUCID 
(see Eq.~\ref{eq:specific-choice-on-x-brh-cent}).
{\revstyle 
In the history, central subhalos can fall into neighboring halos, temporarily 
becoming satellites, before being ejected back to the central phase. 
During their temporary satellite phase, we represent their halo mass by their 
mass right prior to infall, causing a discontinuity in the MAHs of some 
central subhalos, as shown in the right block of 
Fig.~\ref{fig:extension-central-history}. These MAHs exhibit temporary plateaus, 
followed by sudden jumps to higher masses. Many of these infall-ejection events 
are artificial, arising from the bridging effect of the FoF algorithm 
\citep[e.g.,][]{klypinDarkMatterHalos2011}. To mitigate this issue, one can 
simply replace the halo finder with an algorithm that more robustly excludes 
these artificial links \citep[e.g.,][]{klypinParticleMeshCodeCosmological1997,
knollmannAhfAMIGAHALO2009,
planellesASOHFNewAdaptive2010,behrooziROCKSTARPHASESPACETEMPORAL2012,
valles-perezHaloFindingProblem2022}.
}

\subsection{Joint Distribution of Satellite Properties}
\label{ssec:joint-satellite-properties}

\begin{figure*} \centering
    \includegraphics[width=\textwidth]{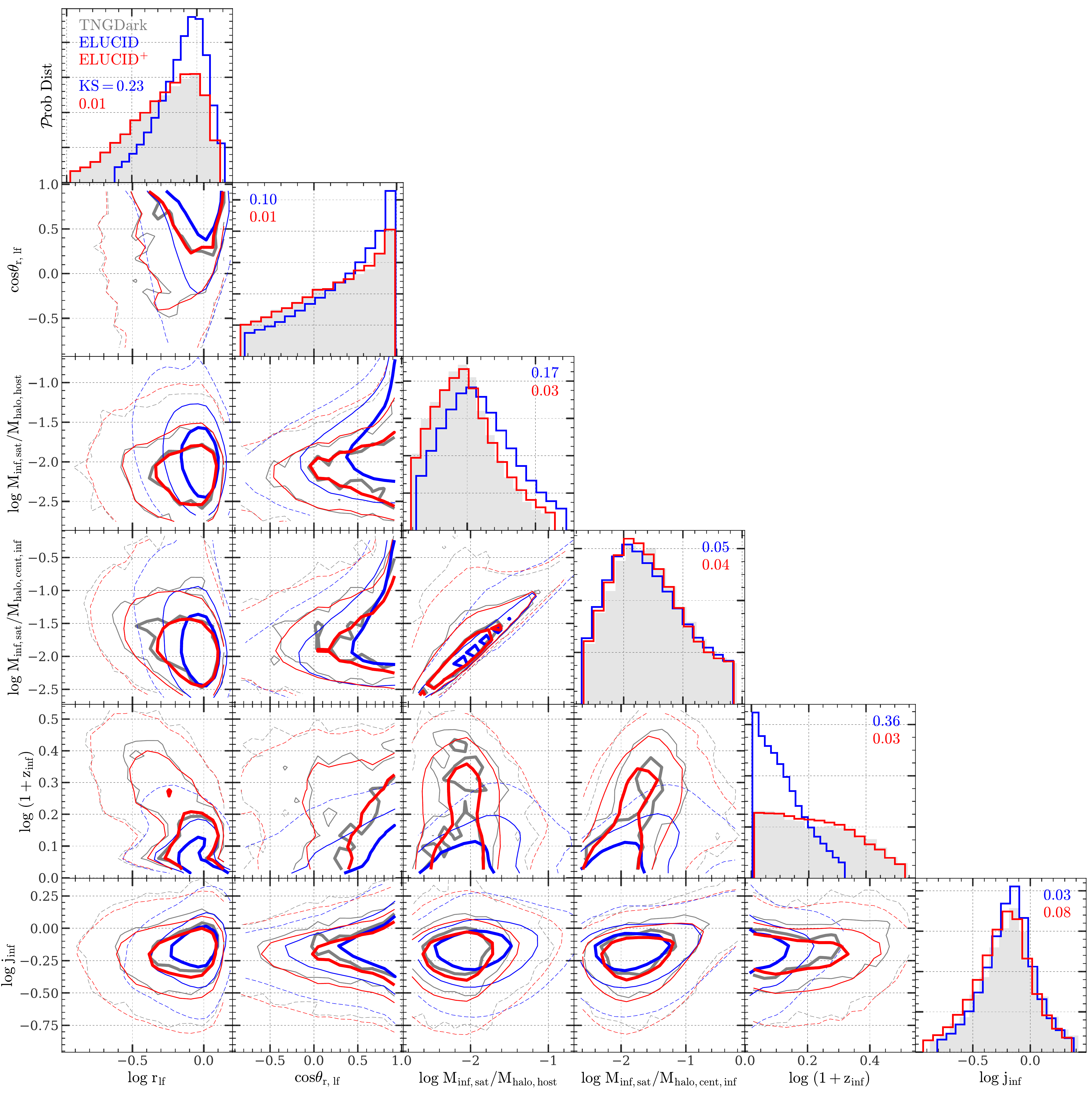}
    \caption{
        Marginal distributions of satellite subhalos in the projected spaces of 
        several properties as indicated by legends of individual axes.
        Satellite subhalos in host halos with 
        $M_{\rm halo, host} \in [10^{12}, 10^{13})\msun$ at $z=0$ are used in 
        the plot. Each diagonal panel shows the 1-D distribution of a property. 
        Each off-diagonal panel shows the 2-D distribution 
        of a pair of properties. The gray, blue, and red histograms or contours
        are the distributions of subhalos in TNGDark, ELUCID and $\rm ELUCID^+$, 
        respectively.
        In each off-diagonal panel, the thick solid, thin solid and dotted lines 
        enclose $30\%$, $60\%$ and $90\%$ of subhalos, respectively.
    }
    \label{fig:extension-2d-hist}
\end{figure*}

As described in \S\ref{sssec:phase-space-assignment}, the goal of the 
phase-space assignment step is to recover the joint distribution 
of a given set of properties, ${\bf x}_{\rm sat}$, of satellite subhalos.
Fig.~\ref{fig:extension-2d-hist} shows the marginal distributions of 
satellite subhalos at $z=0$ in the space of various properties. 
The subhalo properties presented in the figure are the halo-centric radial 
distance $r_{\rm lf}$ and the polar angle $\theta_{\rm r, lf}$,
both defined with respect to the local frame of the host halo, the infall mass 
$M_{\rm inf, sat}$, scaled either by $M_{\rm halo, host}$, the current host halo mass, 
or by $M_{\rm halo, cent, inf}$, the mass of the host halo into which it fell 
at $z_{\rm inf}$, the infall redshift $z_{\rm inf}$, and the 
infall orbital angular momentum $j_{\rm inf}$.
In the 1-D distributions of $r_{\rm lf}$, $\theta_{\rm r, lf}$, 
$M_{\rm inf, sat}/M_{\rm halo, host}$ and $z_{\rm inf}$,
the reference simulation, TNGDark, shows significant
differences from the target simulation, ELUCID.
The difference between the PDF of the two simulations is quantified by the 
Kolmogorov-Smirnov (K-S) statistic, which is larger than $0.1$ 
in each of these four panels. These differences are expected and can 
be interpreted as follows. First, a satellite in ELUCID is 
more likely disrupted artificially in the inner region of its host halo, 
because of the denser environment in that region and the long 
time-integration before arriving there. 
This causes a shift of the PDF towards larger halo-centric distance
as seen in the 1-D panel for $r_{\rm lf}$. The density profiles 
and correlation functions presented in 
Fig.~\ref{fig:extension-prop-dist}, \ref{fig:extension-rsd} and 
\ref{fig:modeled-corrfunc} also show the effects of such incompleteness in ELUCID.
Second, the distribution of satellites resolved in ELUCID tends to align in   
the direction of the COM, as seen from the 1-D PDF of $\cos \theta_{\rm r, lf}$. 
This is partially due to our choice for the polar direction of the spherical 
coordinate system used in Eq.~\ref{eq:spherical-coord-of-x}, and partially due to 
the stronger environmental effect on satellites that are closer to 
${\bf x}_{\rm lf, cent}$, the location of the local potential minimum in the host halo.
Third, a satellite with lower infall mass has shallower local gravitational 
potential to prevent its matter from environmental disruption, 
especially when it approaches the halo center. As a result, the PDF of 
the ratio $M_{\rm inf, sat}/M_{\rm halo, host}$ for ELUCID is shifted towards 
higher values of the ratio. Finally, the shift of the PDF of $z_{\rm inf}$
towards smaller values in ELUCID is a result of the time integration of 
numerical loss. Unlike $M_{\rm inf, sat}/M_{\rm halo, host}$, 
the PDF of $M_{\rm inf, sat}/M_{\rm halo, cent, inf}$ for ELUCID 
shows no significant difference from that for TNGDark. 
This is a coincidence produced by the left-shifted PDF of $z_{\rm inf}$, 
the right-shifted PDF of $M_{\rm inf, sat}/M_{\rm halo, host}$, 
and the positive correlation between $z_{\rm inf}$ and $M_{\rm inf, sat}/M_{\rm halo, cent, inf}$.

The 2-D marginal distributions in Fig.~\ref{fig:extension-2d-hist} present
more demanding tests on satellite properties predicted by ELUCID. The discrepancy  
between ELUCID and TNGDark is even worse in these distributions. Indeed,  
none of these panels shows consistent contours between the two simulations.
This discrepancy indicates that a halo-based galaxy formation model 
applied to ELUCID will not be able to predict reliably the spatial distribution of 
satellite galaxies and the joint distribution between spatial positions and 
other properties of satellite galaxies. 

Thus, extensions of the satellite parts of subhalo merger trees are clearly 
needed by ELUCID. To this end, we separate the difference between ELUCID and TNGDark 
in the joint distribution of satellite properties into two parts. In the first 
part, the difference is the amplitude of the distribution function 
caused by the inadequate number of satellites resolved by ELUCID up to the epoch 
in question. In the second, the difference is the shape of the distribution 
function caused by the dependency of artificial disruption on other satellite properties.
The two parts of the difference are corrected, separately, by two steps of our 
algorithm, the satellite-stage completion (\S\ref{sssec:satellite-stage-completion}) 
and the phase-space assignment (\S\ref{sssec:phase-space-assignment}).

The red histograms and contours labeled as $\rm ELUCID^+$ in 
Fig.~\ref{fig:extension-2d-hist} show the 1-D and 2-D marginal PDFs, 
respectively, of satellite properties after the application of the extension algorithm.
In the 1-D panels, the discrepancy seen between ELUCID and TNGDark is 
completely absent between ${\rm ELUCID}^+$ and TNGDark.   
The K-S statistics between them in all panels are now below $0.1$,
indicating small difference between the two sets of the data after 
the amendment using the extension algorithm. 
The consistency between ELUCID and TNGDark in 2-D distributions is also improved 
significantly after the amendment, as can be seen from the similarity  
in contours between ${\rm ELUCID}^+$ and TNGDark.
Remarkably, in the space of each pair of variables considered 
here, $\rm ELUCID^+$ follows TNGDark closely even in 
their $90\%$ contours. The angular distribution, as represented by 
panels showing pairs that contain $\theta_{\rm r, lf}$, is also well recovered, 
even though we only used the radial distance, $r_{\rm lf}$, as the quantity to match 
in the conditional abundance matching step. This is at least partly because of the 
correlation between $\theta_{\rm r, lf}$ and other conditioning variables.
Although Fig.~\ref{fig:extension-2d-hist} shows only a specific host halo 
mass range, our tests showed that the recovery of the distribution of 
satellite properties in all other halo mass ranges is as good as or even better
than the results presented here. Our tests also showed that the 
algorithm performs equally well for halos identifies 
at $z>0$ (see Fig.~\ref{fig:extension-2d-hist-high-z} for 
an example). At high redshift ($z \gtrsim 4$), the sample size of massive halos 
($M_{\rm halo, host} \gtrsim 10^{12} \msun$) in TNGDark is too small to be robustly 
compared with ELUCID for the joint distribution. In this case,
the split of the full set of satellite properties into conditioning 
and conditioned sets, and the lower bounds we impose on 
$N_{\rm min, cell\ partition}$ and $N_{\rm min, cell\ match}$ in partitioning 
the feature space and matching cells, respectively (see \S\ref{sssec:phase-space-assignment}), 
are the keys to suppressing the cosmic variance and 
to achieving a robust assignment of phase-space coordinates.

\subsection{Summary Statistics of the Subhalo Population}
\label{ssec:summary-statistics}

\begin{figure*} \centering
    \includegraphics[width=0.85\textwidth]{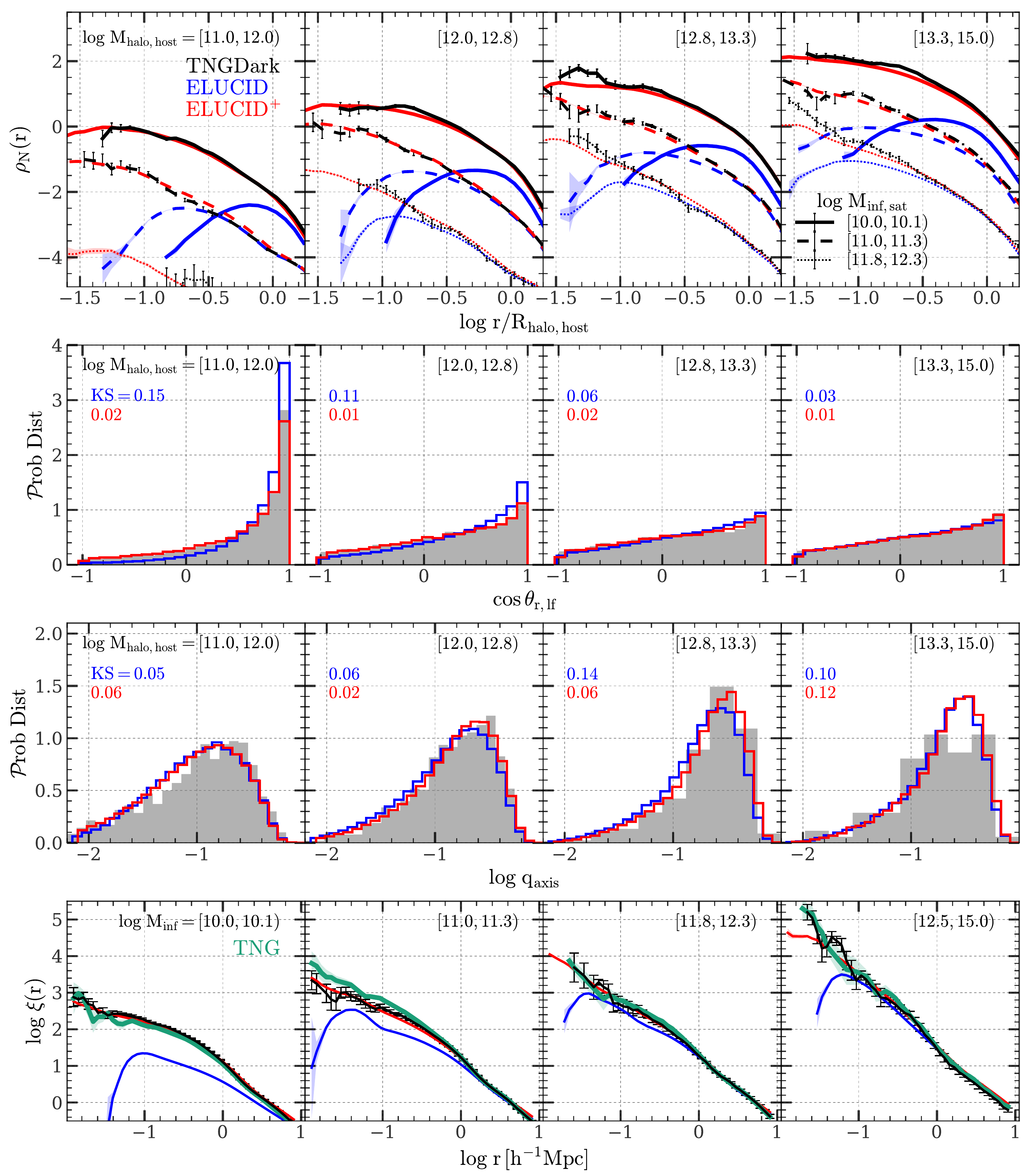}
    \caption{
        Summary statistics for spatial distribution of subhalos at $z=0$. 
        In all panels, black, blue and red symbols are results from TNGDark, ELUCID
        and $\rm ELUCID^+$, respectively. Green curves in the last row 
        are results from the TNG hydro simulation to demonstrate the effects of 
        baryonic processes.
        Errorbars and shaded areas indicate 
        the standard deviations around the corresponding mean values computed 
        from 50 bootstrap samples.
        The first row shows the number density profiles, $\rho_N$, of satellite 
        subhalos in host halos with different masses, $M_{\rm halo,host}/(\msun)$, 
        indicated at the top of panels. 
        For each given halo mass range, satellite subhalos in three different
        infall mass ranges are shown by solid, dashed and dotted lines, respectively,
        and they are shown in an increasing $1 {\rm dex}$ vertical offset for clarity.
        The second row shows the angular distributions of satellite subhalos
        (see Eq.~\ref{eq:spherical-coord-of-x} and texts around it for 
        the definition of the position polar angle $\theta_{r, {\rm lf}}$)
        in host halos with different masses, 
        $M_{\rm halo,host}/(\msun)$, indicated at the top of panels.
        The K-S statistic is computed and indicated in the upper left corner of each panel 
        for the ELUCID (or $\rm ELUCID^+$) distribution with respect to the TNGDark 
        distribution in the same panel.
        The third row shows the distributions of axis ratios of halos with 
        different masses, $M_{\rm halo,host}/(\msun)$, indicated at the top of panels.
        The axis ratio of each halo is computed by using all subhalos (central 
        and satellite) in this halo, weighted by their infall masses.
        The K-S statistics are also indicated in the upper left corner of each panel.
        The fourth row shows the two-point auto-correlation functions of 
        all subhalos (central and satellite) in subsamples with different infall 
        masses, $M_{\rm inf}/(\msun)$, indicated at the top of each panel. 
    }
    \label{fig:extension-prop-dist}
\end{figure*}

The recovery of the joint distribution in space of high-dimensionality 
indicates that other statistical properties of the subhalo population 
are also recovered. For completeness, Fig.~\ref{fig:extension-prop-dist}
shows four statistical measurements that are commonly used in literature.
The first row of Fig.~\ref{fig:extension-prop-dist} shows the number density 
profile, $\rho_N$, as a function of the halo-centric distance $r$ measured 
relative to the central subhalo and scaled by the virial radius, 
$R_{\rm halo,host}$, of the host halo. Results are shown for satellite subhalos 
with different infall masses, $M_{\rm inf, sat}$, and in 
host halos with different masses, $M_{\rm halo, host}$. 
From curves showing TNGDark results, it is clear that the overall amplitude of 
$\rho_N$ is larger for more massive host halos and for less massive satellites.
HOD models \citep[e.g.,][]{jingSpatialCorrelationFunction1998,
berlindHaloOccupationDistribution2002,
guoRedshiftspaceClusteringSDSS2015,
guoModellingGalaxyClustering2016,
yuanAbacusHODHighlyEfficient2022,
qinHIHODHalo2022} are usually parameterized with this assumption. 
The profile decreases monotonically with increasing halo-centric 
distance, which is usually modeled by a double-power-law form, such 
as the NFW \citep{navarroUniversalDensityProfile1997} profile.
With limited resolution, the profiles revealed by the ELUCID simulation, 
as shown by blue curves, lack some of these critical features.
The profiles of ELUCID follow those of TNGDark at large radii, but they 
start to bend down when approaching to inner regions of host halos. 
For satellite subhalos with masses $\sim 10^{10} \msun$, the profiles 
start to deviate from those of TNGDark even at $r\sim R_{\rm halo, host}$. 
Very few subhalos of such mass are present at $r< (1/5) R_{\rm halo, host}$. 
These subhalos have masses too close to the mass resolution limit of ELUCID, 
and are severely affected by numerical artifacts. More massive satellite subhalos 
in ELUCID are more stable against numerical effects,  but they 
are also under-represented  in the inner region of the hosts, 
because their progenitors and structures may not be properly resolved. 
The profiles after extension, marked as ${\rm ELUCID}^+$ and shown by red curves,
are significantly improved. Over the entire ranges of both 
the host halo mass and the satellite infall mass, the extended profiles 
follow tightly those of TNGDark all the way to $r \sim 0.1 R_{\rm halo, host}$.
At $r< 0.1 R_{\rm halo, host}$, the TNGDark profiles become noisy, as seen from the 
large fluctuations and error bars. However, the ${\rm ELUCID}^+$ profiles in 
the innermost regions, $r \sim 10^{-1.5} R_{\rm halo, host}$, are still stable,
owing to the much larger simulation volume and sample size of 
ELUCID in comparison to TNGDark. Note that training of the extension 
algorithm is less demanding on sample size than some statistical 
measures. These results indicate that our extension algorithm is able 
to combine the large volume of the target simulation with the high 
resolution of the reference simulation.

When modeling galaxy formation based on subhalos, the number density profiles 
of satellite galaxies serve as a critical test or calibration for model 
predictions. These profiles provide the ``one-halo'' terms in galaxy 
two-point correlation functions, which can be measured directly from galaxy surveys  
\citep[e.g.,][]{liDistributionStellarMass2009, mengMeasuringGalaxyAbundance2020a}.
With a halo-based group finder \citep[see, e.g.,][]{yangHalobasedGalaxyGroup2005,
yangGalaxyGroupsSDSS2007,wangIdentifyingGalaxyGroups2020}, 
these profiles can also be measured directly by stacking groups of similar 
masses and by properly correcting redshift-space distortions.
Thus, the extension algorithm developed here provides a solid basis to model 
galaxy clustering reliably. 

The second row of Fig.~\ref{fig:extension-prop-dist} shows the angular 
distribution of satellite subhalos in terms of the PDF of the cosine of 
the position angle $\theta_{r, {\rm lf}}$ in host halos of different masses.
The PDFs all have a minimum at $\theta_{r, {\rm lf}} \sim \pi$, 
increase as the polar angle decreases, and reach to a maximum at 
$\theta_{r, {\rm lf}} \sim 0$. This tendency of alignment between the 
halo COM and satellites is an outcome of our definition of the 
spherical coordinate system in the local frame 
(see Eq.~\ref{eq:spherical-coord-of-x}). The alignment is stronger in 
lower-mass host halos and particularly significant in halos with 
$M_{\rm halo, host} < 10^{12} \msun$. This is because lower-mass hosts 
have a smaller number of satellites, which are preferentially 
distributed around the COM. With a limited resolution, ELUCID misses some of the 
satellites, and the missed fraction is more significant for satellites that 
are anti-aligned with the  COM and in less massive hosts. 
By the definition of the polar angle, these anti-aligned satellites are closer to 
the central subhalo on average and have smaller mass to resist numerical noise 
as they approach the potential minimum. After the extension, the PDFs obtained from
${\rm ELUCID}^+$  become indistinguishable from those of TNGDark, indicating that our 
algorithm successfully captures the angular distribution of satellites. 
The K-S statistic, which measures the difference between 
the PDFs of two distributions, is $>0.1$ between ELUCID and TNGDark
for host halos with $M_{\rm halo, host} < 10^{12.8} \msun$
and becomes negligibly small ($ \leqslant 0.02$) between ${\rm ELUCID}^+$
and TNGDark.

The third row of Fig.~\ref{fig:extension-prop-dist} shows the distribution 
of the axis ratio for halos with different masses. Following 
\citet{maccioConcentrationSpinShape2007,chenRelatingStructureDark2020}, we 
use the definition  
\begin{equation}
    q_{\rm axis} = \frac{q_2 + q_3}{2 q_1},
\end{equation}
where $q_1$, $q_2$ and $q_3$ are the three principal axes of 
the inertial ellipsoid computed using all member subhalos 
(central and satellite) weighted by their infall masses. 
So defined, a spherical halo has $q_{\rm axis}=1$
while a needle-shaped halo has $q_{\rm axis}=0$. Compared with 
TNGDark, halos in ELUCID tend to be slightly more elongated, 
as seen in the first three bins of halo masses. 
This is likely caused by the higher odd of disrupting
under-resolved subhalos in inner regions of ELUCID halos
combined with the fact that the distribution of 
satellite subhalos tends to be more spherical in inner regions
of their hosts.
After the extension, the distributions of $q_{\rm axis}$ 
become more like those in TNGDark. This is a result of 
the shape-preserving nature of our algorithm together with the 
recovery of subhalos in the inner regions of host halos.
The K-S values after the extension are reduced for halos with 
$M_{\rm halo, host} \in [10^{12}, 10^{13.3}) \msun$, but 
slightly increased  for halos with $M_{\rm halo, host} < 10^{12} \msun $ 
and $M_{\rm halo, host} \geqslant 10^{13.3} \msun$. The slightly worse 
K-S for the lowest-mass halos is caused by the small number of satellites 
in elongated distribution, as seen from the long tail of the PDF at 
$q_{\rm axis} \sim -2$. For halos in the highest mass bin, the 
TNGDark sample is small, and the reference distribution it provides 
is uncertain, as one can see from the shape of its histogram.  

The position polar angle $\theta_{r, {\rm lf}}$ and the axis ratio 
$q_{\rm axis}$ are two quantities that can be used to describe the 
anisotropic distribution of satellites in host halos. The anisotropic distribution 
of satellites, and its dependence on properties such as color and quenching state, 
have been detected in observations and tested using simulations 
\citep[see, e.g., ][]{ibataVastThinPlane2013, 
yangAlignmentDistributionSatellites2006,brainerdLopsidedSatelliteDistributions2020,
martin-navarroAnisotropicSatelliteGalaxy2021a}. 
Because our extension algorithm is shape-preserving and can 
recover the anisotropic distribution of satellite subhalos, 
halo-based models using the extended trees are expected to 
be able to reproduce the anisotropic distribution, and can be used 
to separate effects produced by the underlying subhalo distribution
from those generated by baryonic processes.

The last row of Fig.~\ref{fig:extension-prop-dist} shows the two-point 
correlation function $\xi(r)$ of subhalos (central and satellite) with 
different infall masses, where $r$ is the separation of subhalo pairs. 
Much like the density profile, the ``one halo'' term of $\xi(r)$ is 
underestimated by ELUCID due to missed subhalos, and its deviation 
from TNGDark becomes more significant in the inner region of host halos. 
Subhalos with larger infall masses in ELUCID are less affected by 
numerical defects, and their $\xi(r)$ follows that of TNGDark better. 
After the extension, the discrepancy is almost completely removed, 
as one can see by comparing ${\rm ELUCID}^+$ with TNGDark. The extension 
allows $\xi(r)$ in the low-resolution target simulation to be 
extended accurately to very small scales. Note also that the 
amended correlation functions (red lines) are much smoother than 
their counterparts in TNGDark (black lines) on scales below 
$0.1 \mpc$, again because of the difference in sample size.
Complementary to the density profile of satellites, the correlation 
function carries additional information about clustering on 
inter-halo scales. Since our extension algorithm does not change 
the ``two-halo'' term, the small differences between ${\rm ELUCID}^+$ 
(or ELUCID) and TNGDark on such scales are due partly to the difference in 
cosmological models adopted in the two simulations 
and partly to cosmic variances in TNGDark. 
These differences can be removed by using identical cosmology 
for both the target and reference simulations, $S$ and $S'$, and 
by taking into account cosmic variances caused by the smaller volume 
of the reference simulation. {\revstyle
In Appendix~\ref{app:completeness-and-convergence}, we assess the performance of 
the extension algorithm by employing a pair of simulations with
identical cosmological parameters and initial condition. Notably, the 
differences in the TPCFs at large radii are effectively removed, as seen
from the darkest orange line and the black line in 
Fig.~\ref{fig:app-tpcf-vol-variants}.
}

It is known that baryonic processes can affect the underlying dark matter
distribution. The baryon component tends to make subhalos more concentrated 
and thus harder to strip by tidal forces in their host halos. 
The difference in the mass that can be retained by a subhalo 
can, in turn, change the orbit of the subhalo. 
However, the distribution of the baryonic component is sensitive 
to the subgrid physics implemented in a hydro simulation, and 
its effects are difficult to quantify in a unified way. 
For example, combining hydrodynamic simulations and subhalo 
abundance matching models, \citet{simhaTestingSubhaloAbundance2012} 
showed that the two-point correlation function of galaxies is affected 
by mass contained in stars, and that the existence of momentum-driven winds 
in hydrodynamic simulations can modify effects of the baryonic component. 
As a test, we compute the two-point correlation functions from TNG, the 
full hydro counterpart of TNGDark, and show the results by green curves 
in the last row of Fig.~\ref{fig:extension-prop-dist}. The difference between 
TNGDark and TNG is much smaller than that caused by numerical resolution 
as measured by the difference between TNGDark and ELUCID, and it is comparable to 
the uncertainty of our extension algorithm as measured by the difference between 
TNGDark and ${\rm ELUCID}^+$. This indicates that our extension algorithm has 
nearly reached the upper limit of the quality provided by the high-resolution 
DMO simulation that does not include baryonic effects.
To include baryonic effects in our modeling, a simple solution is to keep the 
extension algorithm unchanged, but to replace the training simulation, $S'$, with a 
hydro simulation that implements baryonic processes. This solution, however, 
will depend on baryonic processes implemented in and the accuracy of the hydro simulation.


\subsection{Redshift-Space Correlation Functions}
\label{ssec:test-rsd}

\begin{figure*} \centering
    \includegraphics[width=0.82\textwidth]{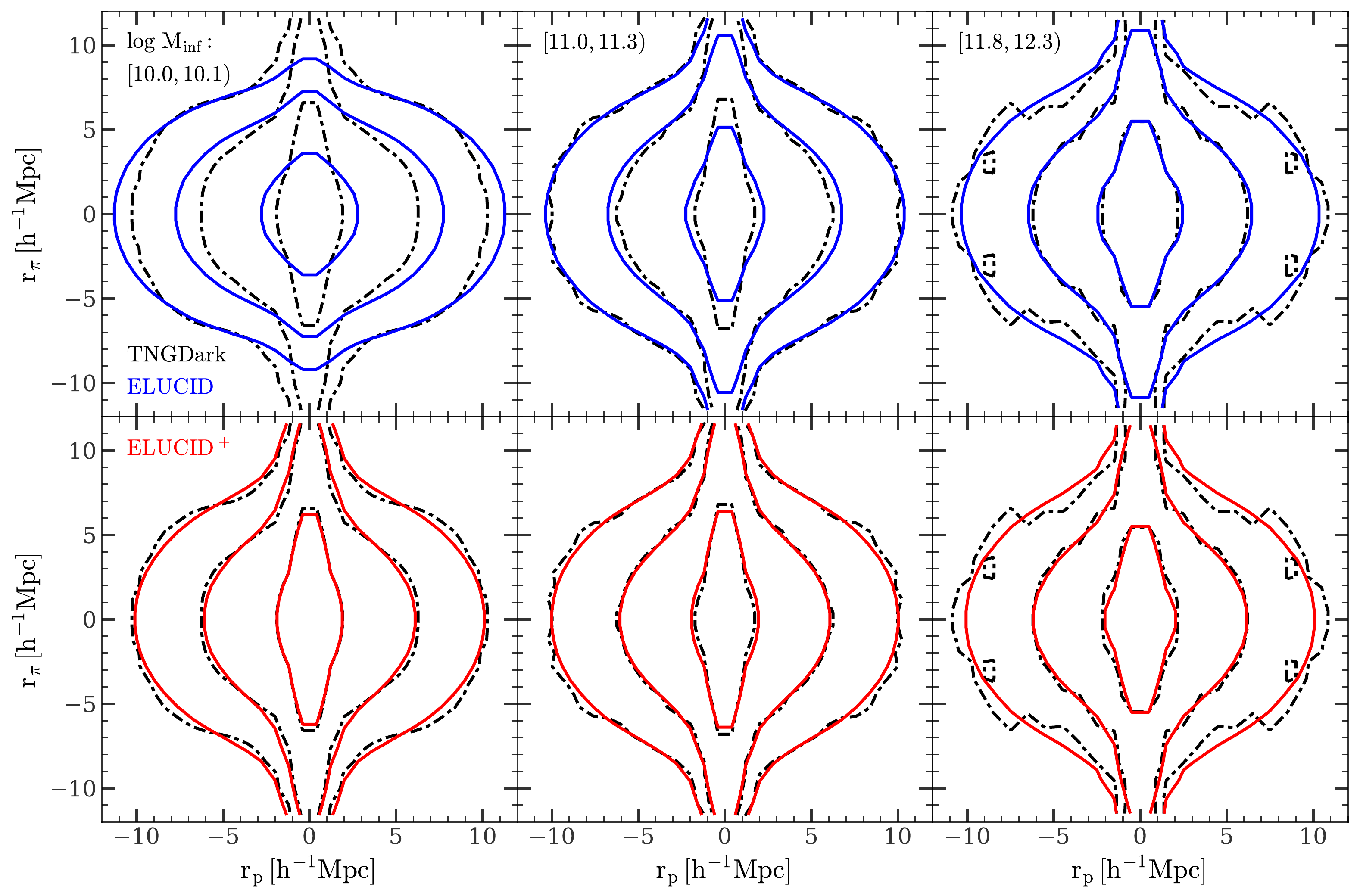}
    \caption{
        Two-dimensional correlation functions $\xi(r_{\rm p}, r_{\rm \pi})$ 
        in redshift space for subhalos (central plus satellite) at $z=0$.
        Three columns show the results for halos with different
        infall masses, $M_{\rm inf} / (\msun)$,
        indicated in the top left corner of the panels in the first 
        row. Black contours in all panels are obtained from TNGDark.
        Red and blue contours in two rows are obtained from 
        ELUCID and its extended version, ${\rm ELUCID}^+$, respectively.
    }
    \label{fig:extension-rsd}
\end{figure*}

Tests presented above are based on positions of subhalos, and it is clearly  
important to check how the extension algorithm performs on modeling peculiar
velocities of subhalos. Accurate phase-space information is critical 
to generating reliable mock galaxy samples that mimic the real observations in redsift 
space. In redshift space, the line-of-sight (LOS) peculiar velocities of subhalos 
distort the pattern of galaxy clustering in space, which is known as the 
Finger of God (FOG) effect on small scales \citep{jacksonFingersGod1972, 
fisherClusteringJyIRAS1994}, and the Kaiser effect on large scales 
\citep{kaiserClusteringRealSpace1987}. The redshift-space distortion 
(RSD) caused by the FOG effect depends on the density and velocity profiles of 
subhalos in their host halos, and so low-resolution simulations may not be able to 
model it accurately.

To see the effect of numerical resolution on RSD, we compute the two-dimensional 
correlation function, $\xi(r_{\rm p}, r_{\rm \pi})$, as a function of the 
projected separation, $r_{\rm p}$, and the LOS separation, $r_{\rm \pi}$,
for pairs of subhalos. The results of TNGDark and ELUCID are shown in the 
first row of Fig.~\ref{fig:extension-rsd} for subhalos (central and satellite) 
of different infall masses. Here, we use the $z=0$ snapshot 
and choose the $z$-axis of the simulation box as the LOS direction.
For low-mass subhalos with $M_{\inf}\sim 10^{10} \msun$, the FOG effect is 
severely suppressed in ELUCID, as seen from the less elongated contours of 
$\xi(r_{\rm p}, r_{\rm \pi})$. This is expected, because a lower-mass 
satellite has a larger probability to be artificially destroyed in ELUCID due to 
the limited resolution, as seen from the first row of Fig.~\ref{fig:extension-prop-dist}. 
In contrast, subhalos with higher masses, such as those with $M_{\rm inf} \sim 10^{12} \msun$,
are less likely to be missed, and their RSD patterns in ELUCID follow 
better those in TNGDark.  

The two-dimensional correlation function of the extended population with 
reassigned phase-space coordinates are shown in the second row of 
Fig.~\ref{fig:extension-rsd}. Comparing with the original ELUCID results,
we see that the discrepancy with TNGDark for low-mass subhalos 
is completely removed and that the contours of 
$\xi(r_{\rm p}, r_{\rm \pi})$ from ${\rm ELUCID}^+$ match well 
with their TNGDark counterparts. For high-mass subhalos, the improvement
is still evident but less remarkable, because of the smaller 
difference between ELUCID and TNGDark to start with.
Overall, ${\rm ELUCID}^+$ results match those of TNGDark very 
well. Contours of ${\rm ELUCID}^+$ are significantly smoother, 
again because of the significantly larger simulation volume of ELUCID.

\subsection{Subhalos in Individual Halos}
\label{ssec:individual-halos}

\begin{figure*} \centering
    \includegraphics[width=\textwidth]{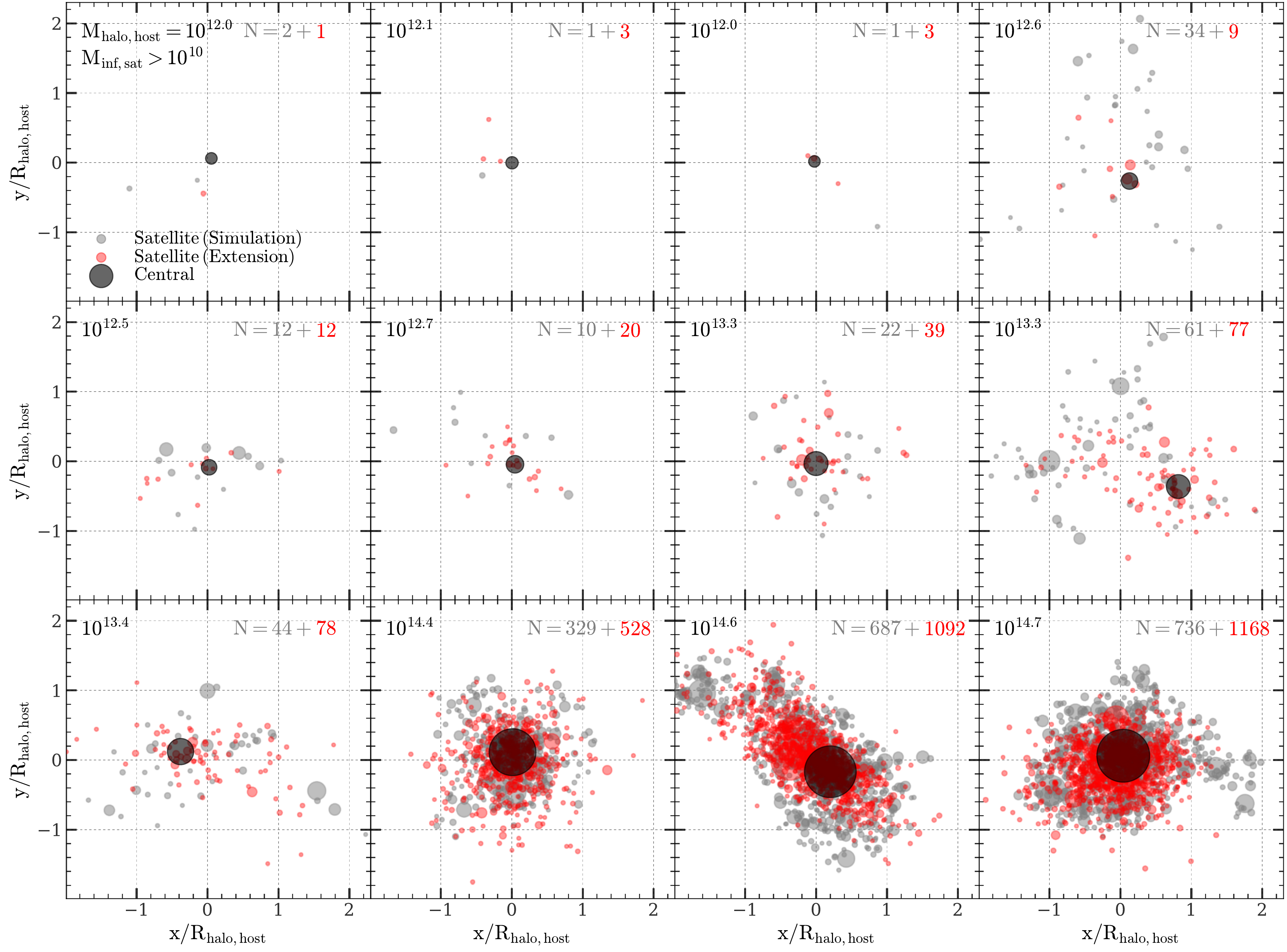}
    \caption{
        Subhalo distributions in the real space of several example halos in ELUCID. 
        Each panel shows the subhalos in one host halo whose mass, $M_{\rm halo,host}/(\msun)$, 
        is indicated in the top left corner of that panel. 
        Black, gray and red dots represent central subhalo, 
        satellite subhalos resolved by ELUCID, and satellite subhalos generated by the extension algorithm, 
        respectively.
        All subhalos with mass greater than $10^{10} \msun$ 
        are shown. Radius of a dot is proportional to the square root of 
        the subhalo infall mass, $M_{\rm inf}$.
        The numbers of simulated and extended satellite subhalos are separately 
        indicated in the upper right corner of the panel. The origin of each panel is 
        the center of mass of the host halo, computed by using all the particles linked
        to it.
    }
    \label{fig:extension-spatial-points}
\end{figure*}

As a visual inspection, Fig.~\ref{fig:extension-spatial-points} shows 
some examples of the spatial distributions of satellites in individual 
halos randomly picked from the population of $M_{\rm halo, host} \geqslant 10^{12} \msun$. 
The central, simulated, and extended subhalos are presented by symbols of 
different colors. The numbers of simulated and extended satellites are 
listed in each panel. The number of satellites with infall mass above $10^{10} \msun$ 
is usually smaller than 10 for halos of $~10^{12} \msun$, less than 
100 for halos with mass $~10^{13} \msun$, and over 1000 for the largest
halos with mass $> 10^{14.5} \msun$. Over the entire range of host halo mass,
a non-negligible fraction of satellites is not properly resolved by the 
ELUCID. The missed satellites are comparable to the simulated ones in 
their total number, but are usually less massive, as seen from the smaller symbol  
sizes. This is consistent with the number density profiles shown in the first row of 
Fig.~\ref{fig:extension-prop-dist}.
Note that phase-space coordinates of most of the simulated satellites are preserved
and assignment is made mainly for low-mass satellites that are 
not properly resolved by ELUCID. This is an outcome of the ``self-consistency'' strategy 
in the conditional abundance matching step (see Eq.~\ref{eq:cond-matching-radial-criterion} 
and the texts around it) intended to preserve as much as possible 
the phase-space information contained in the original simulation.

As one can see, halos are diverse in shape: some are quite round, such as 
those in the the 7th and 10th panels, some are 
elongated, as shown in the 9th and 11th panels. 
Low-mass halos with $M_{\rm halo, host} \leqslant 10^{12} \msun$ 
have too few members to exhibit any regular structure, and 
this is the reason why we use dark matter particles to trace shapes of 
halos in ELUCID for the local frame transformation (see Eq.~\ref{eq:eigen-axes} 
and \ref{eq:local-frame-transformation}). Most of the halos are in 
relaxted states, as indicated by the small offset between the COM and the central subhalo.
The halo in the 8th panel has three massive structures, resulting in a large 
offset between the COM and the central subhalo. The existence of 
un-relaxed systems like this one motivates our choice of the 
reference direction in defining the spherical coordinate system 
(Eq.~\ref{eq:spherical-coord-of-x}) and the inclusion of the relaxation 
indicator $r_{\rm lf, com}$ in the set of conditioning variables 
${\bf x}_{\rm sat, complete}$ for the phase-space assignment 
(Eq.~\ref{eq:conditioning-and-conditioned-vars}).
Taking account of halo shape and relaxation state, the extended satellite 
population follows well the anisotropic distribution around the central 
subhalo, preserving the shapes of halos in all cases shown in  
Fig.~\ref{fig:extension-spatial-points}.

\subsection{Performance on Halo-Based Galaxy Modeling}
\label{ssec:halo-based-model}

\begin{figure*} \centering
    \includegraphics[width=0.85\textwidth]{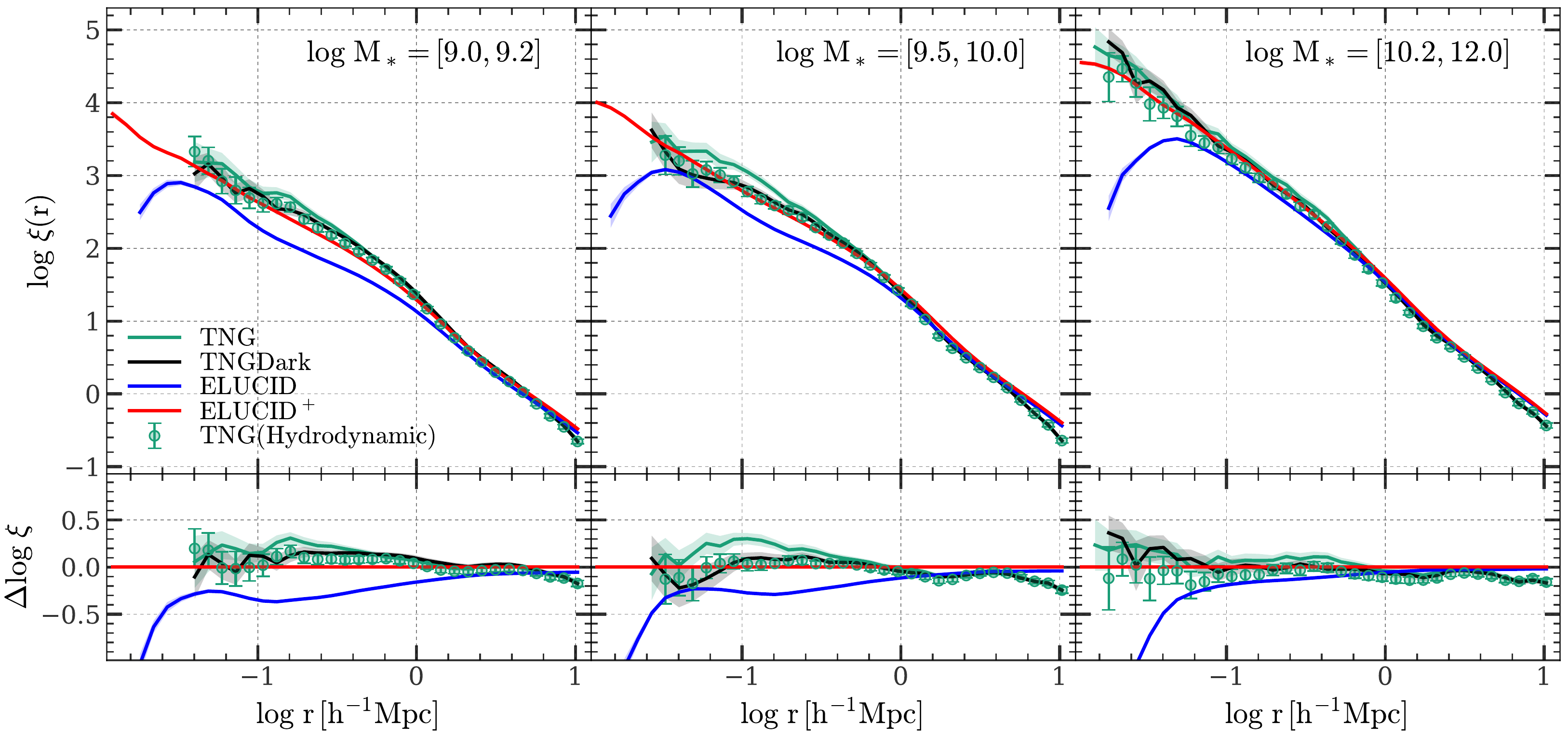}
    \caption{
        Real space correlation functions of $z=0$ galaxies with different stellar masses 
        indicated in the top right corner of each panel. These galaxies are 
        obtained by applying a halo-based empirical model adapter, 
        MAHGIC, to four versions of subhalo merger trees.
        Green, black, blue and red solid curves are the results of the empirical 
        model implemented with subhalo merger trees in TNG, TNGDark, 
        ELUCID and $\rm ELUCID^+$, respectively.
        Green dots are results of the simulated galaxies from the TNG simulation.
        The first row shows the correlation
        functions $\xi(r)$, and the second row shows the difference of
        each correlation function with respect to $\rm ELUCID^{+}$ in the same range 
        of stellar mass. The shaded areas and error bars represent the standard 
        deviations computed from 50 boostrap samples. 
    }
    \label{fig:modeled-corrfunc}
\end{figure*}

The tests presented above verifies that the extended subhalo 
merger trees recover well the joint distribution of various 
subhalo properties, including the infall properties (redshift $z_{\rm inf}$, 
mass $M_{\rm inf, sat}$, mass of host halo $M_{\rm halo, cent, inf}$, and orbital 
angular momentum $j_{\rm inf}$ relative to its central subhalo), the current 
properties (host halo mass $M_{\rm halo, host}$ and phase-space coordinates 
${\bf r}$ and ${\bf v}$). Because these properties are often used as the 
building blocks of halo-based galaxy formation models, the extended subhalo 
merger trees thus form a statistically robust and unbiased basis 
to model galaxies. By so doing, these models automatically take advantages of the large 
simulation volume given by the parent (target) simulation and the well-resolved subhalo 
population given by the extension.

As an example, Fig.~\ref{fig:modeled-corrfunc} shows the two-point correlation 
function of galaxies generated by a halo-based empirical model adapter, MAHGIC, 
developed by \citet{chenMAHGICModelAdapter2021}. The adapter uses a 
flexible pipeline, consisting of dimension transformations and non-linear 
regressors, to map subhalo merger trees to galaxies. The structure and 
parameters of the pipeline can be trained by subhalos and galaxies from 
hydrodynamic simulations or by summary statistics of galaxies from observations 
(Chen et al., in preparation). The pipeline can thus be adapted to a wide set of 
halo-galaxy inter-connections underlying the training data. 
Here, we choose the version of this model that is trained by subhalos and galaxies 
from TNG, and we implement it to different versions of subhalo merger trees. 
Because these implementations share the same halo-galaxy mapping, we are able 
to quantify the difference in the predicted galaxy population caused by the 
difference in the subhalo population between the two implementations.  
The results of two-point correlation function are shown by 
colored curves in Fig.~\ref{fig:modeled-corrfunc} for modeled galaxies 
of different stellar masses at $z=0$. For comparison, we also plot the 
correlation functions of galaxies obtained from the TNG simulation selected in the 
same redshift and stellar mass ranges. The results can be interpreted as follows. 
First, The correlation functions of modeled galaxies based on TNG 
subhalos (green curves) are moderately different from those simulated by TNG (green dots).
This simply indicates that the empirical model, implemented to trees that are 
consistent with the constraining data, is both stable and accurate in  
reproducing galaxy clustering statistics.
The difference in the correlation function is negligible on $r > 1 \mpc$ 
for all galaxies, and smaller than $\sim 0.3\, {\rm dex}$ for galaxies of 
intermediate stellar mass ($\sim 10^{10} \msun$) in inner regions of host halos. 
Because such stellar masses are close to the characteristic mass of the stellar 
mass function, so that different feedback processes may affect the formation and evolution 
of these galaxies, an accurate prediction of their stellar masses is challenging. 
Second, the correlation functions of modeled galaxies based on TNGDark 
(black curves) do not show any bias in comparison with those given by TNG. 
This is a synergistic result of the facts that
baryonic components have only small effect on the correlation functions 
of subhalos, as seen in the fourth row of Fig.~\ref{fig:extension-prop-dist},
and that the empirical model is capable of reproducing galaxy clustering 
statistics from reliable subhalo merger trees.
Third, the correlation functions of modeled galaxies based on ELUCID 
(blue curves) are significantly underestimated on small scales and overestimated 
on large scales, in comparison with TNG results. This is again expected and follows 
from the behavior of correlation functions of subhalos shown in the fourth row of 
Fig.~\ref{fig:extension-prop-dist}. Finally, with the amended subhalo merger 
trees in ${\rm ELUCID}^+$ (red curves), the small-scale bias in 
the galaxy correlation functions is largely reduced. The difference 
with TNG is reduced to $\lesssim 0.2\, {\rm dex}$, comparable to the uncertainty from 
the empirical model. Thus, with a combination of robust statistics from ELUCID 
and the high resolution from TNGDark, the amended correlation functions 
can be measured reliably over the entire range of $r \geqslant 10^{-2} \mpc$. 
Note that on scales $r \lesssim 10^{-1.5}\mpc$, TNG-based correlation functions 
are too noisy to be displayed.

Since our algorithm assigns various properties to the extended subhalos in 
a statistically unbiased manner, (sub)halo-based galaxy models that use
secondary subhalo properties (in addition to mass parameters)  
as inputs to predict galaxy properties can be applied to the extended 
subhalo merger trees. For example, age-matching techniques 
\citep[][]{hearinDarkSideGalaxy2013,hearinDarkSideGalaxy2014,
mengMeasuringGalaxyAbundance2020a,wangRelatingGalaxiesDifferent2023}
rely on mass and formation time of individual subhalos as the main and 
secondary matching properties, respectively, to assign  
galaxies with stellar mass and color (or star formation rate). 
These models can capitalize on the secondary properties 
of subhalos in our extended trees to make detailed 
predictions of the galaxy population using large N-body simulations.   
We will come back to this in a forthcoming paper. 

\section{Summary and Discussion}
\label{sec:summary}

We develop a novel algorithm to extend subhalo merger 
trees in a low-resolution simulation by conditionally matching them with 
trees and subhalos obtained in a high-resolution simulation. The extension 
enables a large DMO simulation to obtain a large set of trees for statistical 
studies and, at the same time, to have sufficient resolution for reliable 
implementations of (sub)halo-based models of galaxy formation. 
The algorithm can be summarized briefly as follows:
\benum
\item For a target low-resolution DMO simulation carried out in a large volume, 
we find a high-resolution simulation run with a similar cosmology. 
We build subhalo merger trees for both of them using a similar method.
\item 
We extend the resolution of each target tree in the low-resolution simulation
by the four steps outlined \S\ref{ssec:outline-algorithm} and detailed 
in \S\ref{ssec:extension-algorithm}. The first step is to separate  
each tree into disjoint branches. Each branch has a central stage, in which the subhalo is a central, 
and a satellite stage, in which the subhalo is a satellite in a host halo.
The second is the central-stage completion of branches, where assembly histories of 
central subhalos are extended to high $z$. 
The third is the satellite-stage completion of branches, in which  
the lifetimes of satellite subhalos are extended beyond the numerical 
disruptions in the target simulation. 
The fourth step is to assign phase-space coordinates  
(positions and velocities) to satellite subhalos through
abundance matching conditioned on cells found by a CART tree.
\item 
We make specific choices of quantities and parameters for the extension algorithm, 
based on the data available and target properties to be recovered, 
and we instantiate each of the above four steps using these choices.
\eenum

We present various tests on the algorithm by extending subhalo merger trees in 
ELUCID, a low-resolution target simulation of large volume, 
with trees from TNGDark, a high-resolution reference simulation run in a smaller box. 
We compare the extended trees with the original ones of ELUCID and with    
those from TNGDark. We also check how well the properties of individual 
subhalos and subhalo populations are recovered by our algorithm.  
Our main conclusions are summarized as follows:
\benum
\item 
Satellite subhalos created by the extension at $z=0$ dominate the low-mass 
end of the halo mass function near the resolution limit ($\sim 10^{10} \msun$
for ELUCID), and have a moderate effect, $\sim 0.15\, {\rm dex}$, 
at the high-mass end (see Fig.~\ref{fig:extension-hmf} and \S\ref{sec:method}).
\item 
The MAHs of individual central subhalos are extended smoothly  
to high redshift until the resolution limit of the reference simulation is reached. 
The average of the extended MAHs over all central subhalos 
matches accurately that of the reference simulation (see 
Fig.~\ref{fig:extension-central-history} and \S\ref{ssec:test-mah}).  
Thus, the extended subhalo mergers trees are not only unbiased, but 
also cover early histories of their formation.  
\item 
The joint distribution of various satellite properties, such as 
phase-space coordinates and infall properties, is statistically 
recovered by the extension. Critical summary statistics, such as 
density profiles and angular distributions of satellites, the shape 
distributions of host halos, the one-dimensional and two-dimensional 
two-point correlation functions, are also improved significantly, 
especially for low-mass subhalos 
(see Fig.~\ref{fig:extension-2d-hist}, \ref{fig:extension-prop-dist} 
and \ref{fig:extension-rsd};
\S\S\ref{ssec:joint-satellite-properties}, \ref{ssec:summary-statistics} 
and \ref{ssec:test-rsd}). 
\item
The ``shape-preserving'' and ``self-consistent'' schemes used in the algorithm
can keep the information from the original target simulation to a maximal extent. 
Thus, the extended subhalos have properties and distributions that 
are consistent with resolved properties in the target simulation, 
such as orientations and shapes of host halos, and phase-space distribution 
of subhalos (see Fig.~\ref{fig:extension-spatial-points} and 
\S\ref{ssec:individual-halos}). 
\item 
With the extended subhalos, a halo-based model of galaxy formation 
can produce satellite galaxies that are statistically unbiased and 
maximally compliant to the original target simulation (see  
examples in Fig.~\ref{fig:modeled-corrfunc} and \S\ref{ssec:halo-based-model}).
\eenum

{\revstyle
The performance of the extension algorithm depends on the resolution of the 
simulation pair and the desired summary statistics. To determine the reference 
simulation requirements and the extension's limitations, a completeness and convergence 
test is necessary (see Appendix~\ref{app:completeness-and-convergence}). 
Furthermore, the simulation pair should have identical cosmology to eliminate 
any differences in the simulated population that are not changed by the extension. 
In case of an application involving variable cosmology of the target simulation, 
the rescaling techniques proposed by \citet{anguloOneSimulationFit2009} can be 
employed to adapt the reference simulation to the target cosmologies before 
applying the extension algorithm.
}

In comparison with other extension methods listed in \S\ref{sec:intro}, 
our extension method for the central MAHs is more precise than the EPS-based 
method \citep{chenELUCIDVICosmic2019,yungSemianalyticForecastsJWST2022,
yungSemianalyticForecastsRoman2022}, retains more information 
from the original simulation than the brute-force joining of 
extensions to root subhalos 
\citep{yungSemianalyticForecastsJWST2022,yungSemianalyticForecastsRoman2022},
and produces smoother transition at the joint redshifts than 
the joining method that does not take into account subhalo formation time 
\citep{chenELUCIDVICosmic2019}. For the extension of satellite subhalos, 
our method produces phase-space coordinates that are correlated 
with subhalo- and host-halo properties, such as infall 
properties, current host halo mass and shape. This allows halo-based 
galaxy formation models to have more input from the halo population  
than methods based on simple assumptions of density and velocity profiles 
\citep{yuanCanAssemblyBias2020,yuanAbacusHODHighlyEfficient2022,
yuanFullForwardModel2022}. Our method is also more physically self-consistent 
than particle-based assignments of phase-space 
coordinates \citep{coleHierarchicalGalaxyFormation2000,
laceyUnifiedMultiwavelengthModel2016,baughGalaxyFormationPlanck2019,
henriquesGalaxyFormationPlanck2015,henriquesLGALAXIES2020Spatially2020}.

The particle-based assignment of phase-space coordinates, however, has 
an advantage that our algorithm does not: it can assign orbits to 
satellites. A limitation of our current method is that it does not track 
orbits for the extended satellites, as our conditional abundance matching 
is performed separately for different snapshots.
A possible solution is to perform the conditional abundance matching 
for whole merger trees instead of for individual subhalos. Unfortunately, 
tree properties are complex, and it is unclear which and in which 
order tree properties should be used in the matching 
\citep[see][for an example of defining a single entropy parameter to characterize a tree]{obreschkowCharacterisingStructureHalo2020}. Thus, tree-based matching 
needs substantially more training data from the reference simulation, 
and may eventually lose its appeal of using high-resolution simulations 
of small volumes as training data. Another solution is to use analytical 
approximations \citep[see, e.g., the orbit-based semi-analytical methods developed by][]{zentnerPhysicsGalaxyClustering2007,jiangSatGenSemianalyticalSatellite2021}
to generate orbits. For the method to work properly, it should not only
retain information from the target simulation to ensure self-consistency, 
but also be able to reproduce joint distributions of satellite properties.
Related tests are yet to be done. 
We will explore these possibilities in the future.

\section*{Acknowledgements}

YC is supported by China Postdoctoral Science Foundation 
(grant No. 2022TQ0329). 
HYW is supported by the National Natural Science Foundation 
of China (Nos. 12192224 and 11890693) and CAS Project for Young Scientists in
Basic Research ({\revstyle grant No. YSBR-062}). 
XY is supported by the National Natural Science Foundation of China
(Nos. 11833005, 11890692).
The authors acknowledge the Tsinghua Astrophysics High-Performance Computing 
platform at Tsinghua University and Supercomputer Center of University 
of Science and Technology of China for providing computational and data storage 
resources that have contributed to the research results reported in this 
paper. YC thanks Wentao Luo for useful discusions. 

\section*{Data availability}

Open source code for the extension algorithm is available in 
Github\footnote{\url{https://github.com/ChenYangyao/merger-tree-extension}}. 
The computation in this paper is supported by the HPC toolkit 
\specialname[Hipp] \citep{2023ascl.soft01030C}
\footnote{\url{https://github.com/ChenYangyao/hipp}}.
Data from the ELUCID project are available at the project 
website\footnote{\url{https://www.elucid-project.com}}. 
All data from the TNG simulation are available at 
the TNG website\footnote{\url{https://www.tng-project.org}}.

\bibliographystyle{mnras}
\bibliography{references}

\appendix
\section{Results at Different Redshifts}
\label{app:high-z}

\begin{figure*} \centering
    \includegraphics[width=0.6\textwidth]{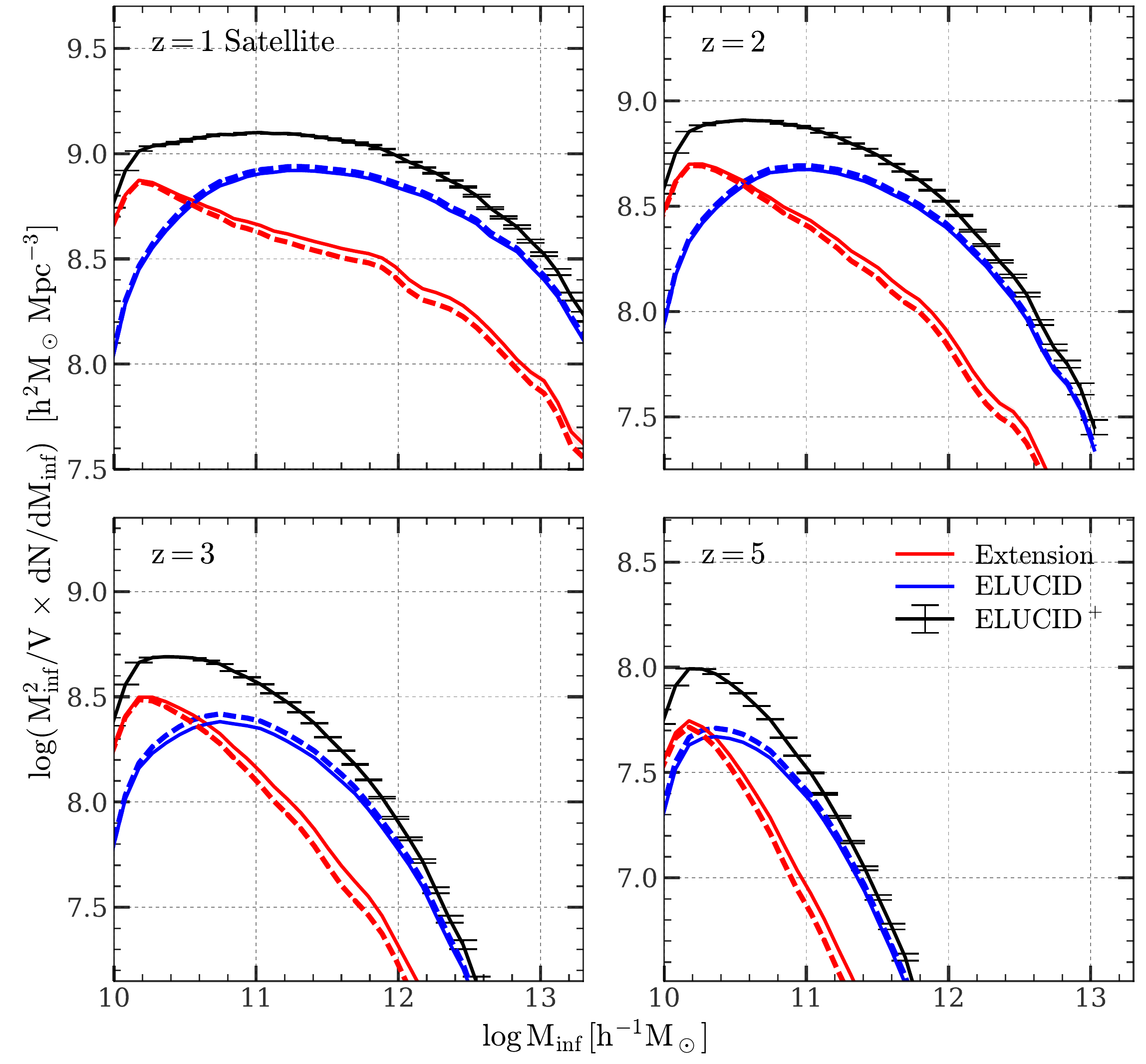}
    \caption{
        Infall mass functions of satellite subhalos in ELUCID. 
        This figure is the same as Fig.~\ref{fig:extension-hmf}, but for satellite 
        subhalos selected at $z=1$, $2$, $3$ and $5$, respectively.}
    \label{fig:extension-hmf-zs}
\end{figure*}

\begin{figure*} \centering
    \includegraphics[width=\textwidth]{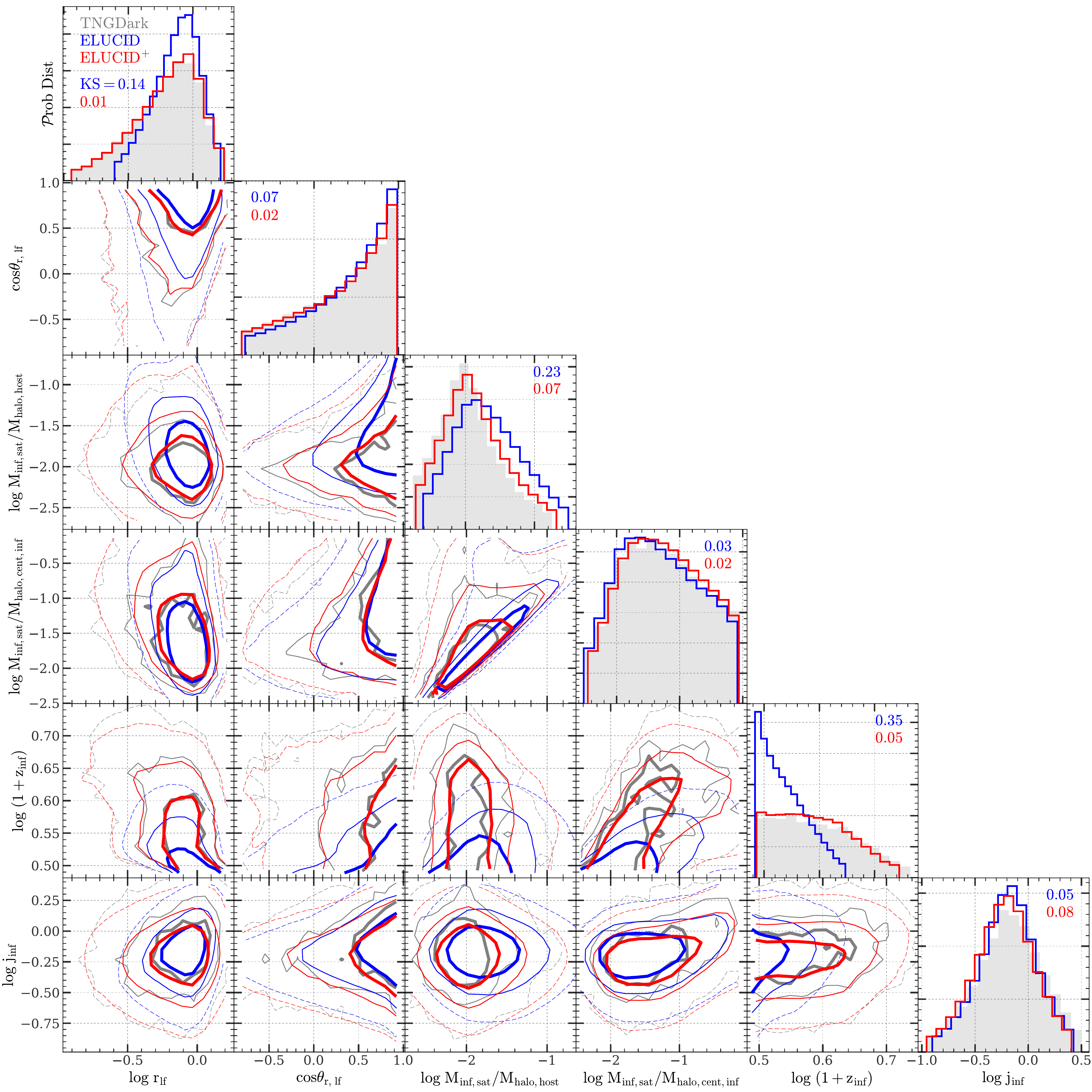}
    \caption{Marginal distributions of satellite subhalos in the projected 
        spaces of properties as indicated by legends of individual axes.
        This figure is the same as Fig.~\ref{fig:extension-2d-hist}, but for 
        satellite subhalos at $z=2$. }
    \label{fig:extension-2d-hist-high-z}
\end{figure*}

In this appendix, we demonstrate the performance of our tree extension algorithm at $z>0$. 
Here, we use the same simulations, $S={\rm ELUCID}$ and $S'={\rm TNGDark}$, 
and adopt the same choices of subhalo properties and algorithm parameters, as 
those specified in \S\ref{ssec:specific-elucid-and-tngdark} and summarized 
in Table~\ref{tab:specific-choices}.

Fig.~\ref{fig:extension-hmf-zs} shows the infall mass functions for satellite
subhalos at four different redshifts. Similar to the results at $z=0$
shown in Fig.~\ref{fig:extension-hmf}, the 
extended subhalos dominate the low-mass end ($M_{\rm inf} \sim 10^{10} \msun$).
The amended mass function is about $0.6\ {\rm dex}$ ($0.2\ {\rm dex}$) 
larger than the original one at $z=1$ ($z = 5$), indicating again 
the importance of the extended population in subhalo statistics. 
At higher infall mass ($M_{\rm inf} > 10^{10.5} \msun$), the simulated 
subhalo outnumbers the extended ones, but the extension still has 
noticeable effects on the mass function.

Fig.~\ref{fig:extension-2d-hist-high-z} shows the marginal distributions
of satellite subhalos in the space of various properties at $z=2$. Similar
to the results of $z=0$ shown in Fig.~\ref{fig:extension-2d-hist}, subhalos 
simulated by ELUCID are significantly different from those by the reference
simulation, TNGDark, in the one-dimensional marginal distributions of 
$r_{\rm lf}$, $\theta_{\rm r,lf}$, $M_{\rm inf, sat}/M_{\rm halo, host}$ 
and $z_{\rm inf}$, as well as in all the two-dimensional marginal distributions. 
This again indicates the incompleteness of the satellite population in ELUCID 
and that the incompleteness depends on subhalo properties. Distributions of 
the amended population in $\rm ELUCID^{+}$ match almost perfectly those 
in TNGDark, as seen from a comparison between the results shown by the red 
and grey colors. The K-S statistics of the 1-D marginal distributions 
between $\rm ELUCID^{+}$ and TNGDark are all less than $0.1$, indicating a 
good match. All these again verify the reliability and precision of our
extension algorithm.

\section{Completeness and Convergence of the Extension}
\label{app:completeness-and-convergence}

\begin{figure*} \centering
    \includegraphics[width=0.6\textwidth]{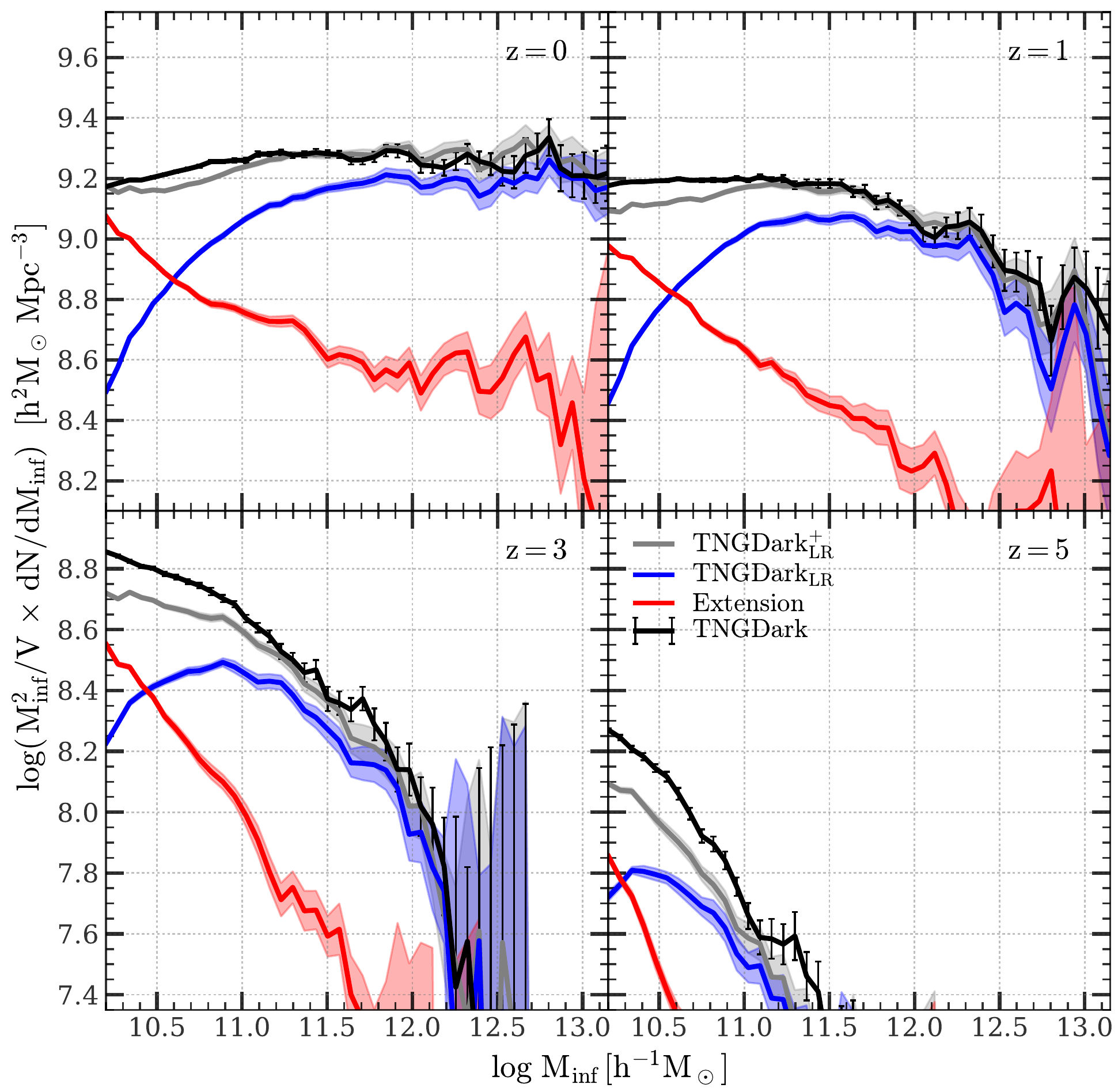}
    \caption{\revstyle
        Infall mass functions of satellite subhalos in TNGDark (black line), 
        $\rm TNGDark_{\rm LR}$ (blue line), and the amended version, 
        $\rm TNGDark_{\rm LR}^+$ (gray line), at redshifts $z=0$, $1$, $3$, 
        and $5$. The red lines in the graph indicate the subhalos created 
        through the extension. All other details are consistent with what was 
        presented in Fig.~\ref{fig:extension-hmf}. }
    \label{fig:app_mfs_completeness_result}
\end{figure*}

\begin{figure} \centering
    \includegraphics[width=0.8\columnwidth]{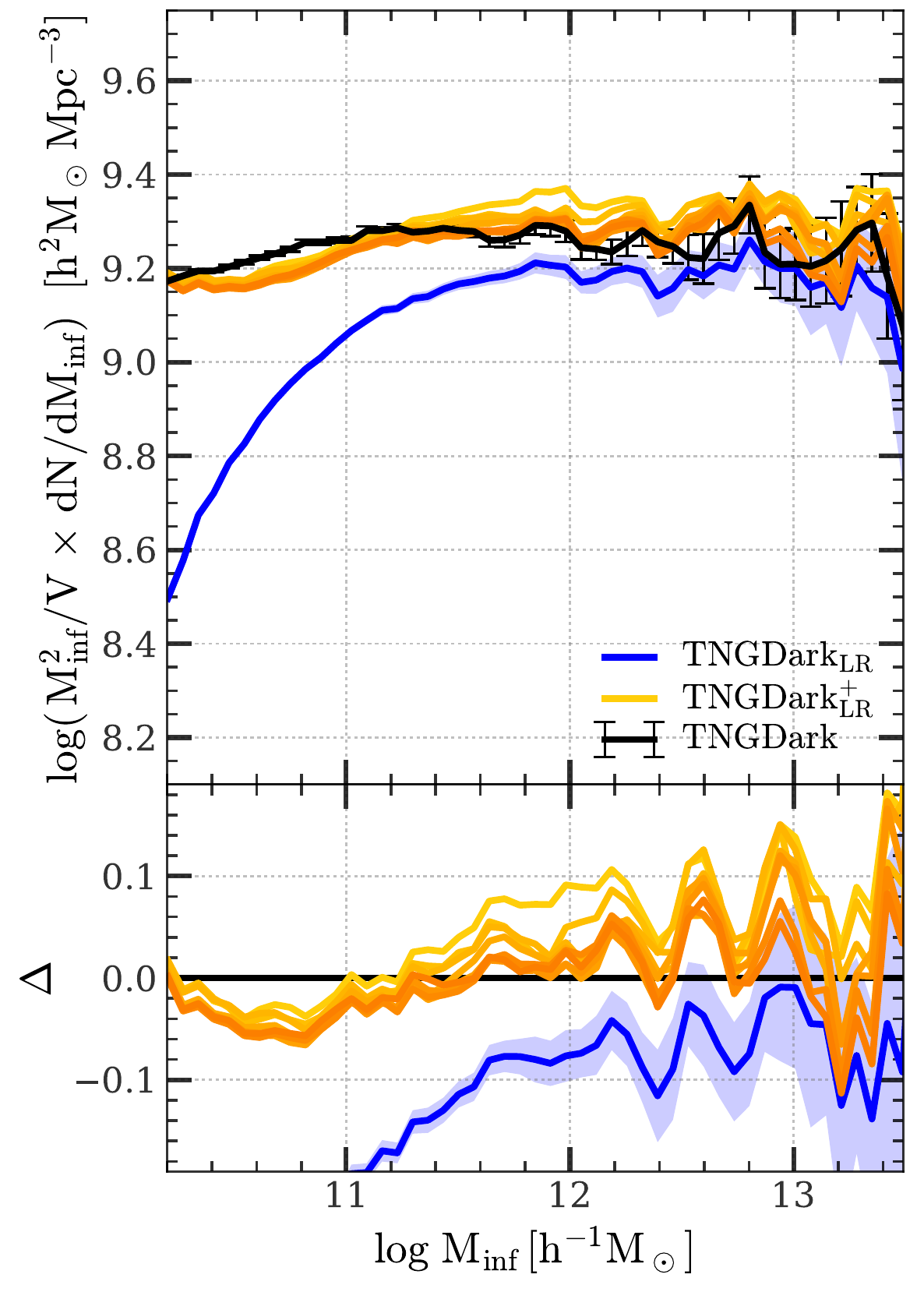}
    \caption{\revstyle
        The same as Fig.~\ref{fig:app_mfs_completeness_result} but here we show 
        the infall mass functions of satellite subhalos at $z=0$ 
        extended using various 
        subvolumes of the reference simulation. The results are represented 
        by orange lines, from the lightest to darkest shade, 
        corresponding to subvolumes of $2\%$, $4\%$, $8\%$, $15\%$, $30\%$, $51\%$, 
        $81\%$, and $100\%$ of the reference simulation's volumes, respectively. }
    \label{fig:app-mf-vol-variants}
\end{figure}

\begin{figure*} \centering
    \includegraphics[width=0.6\textwidth]{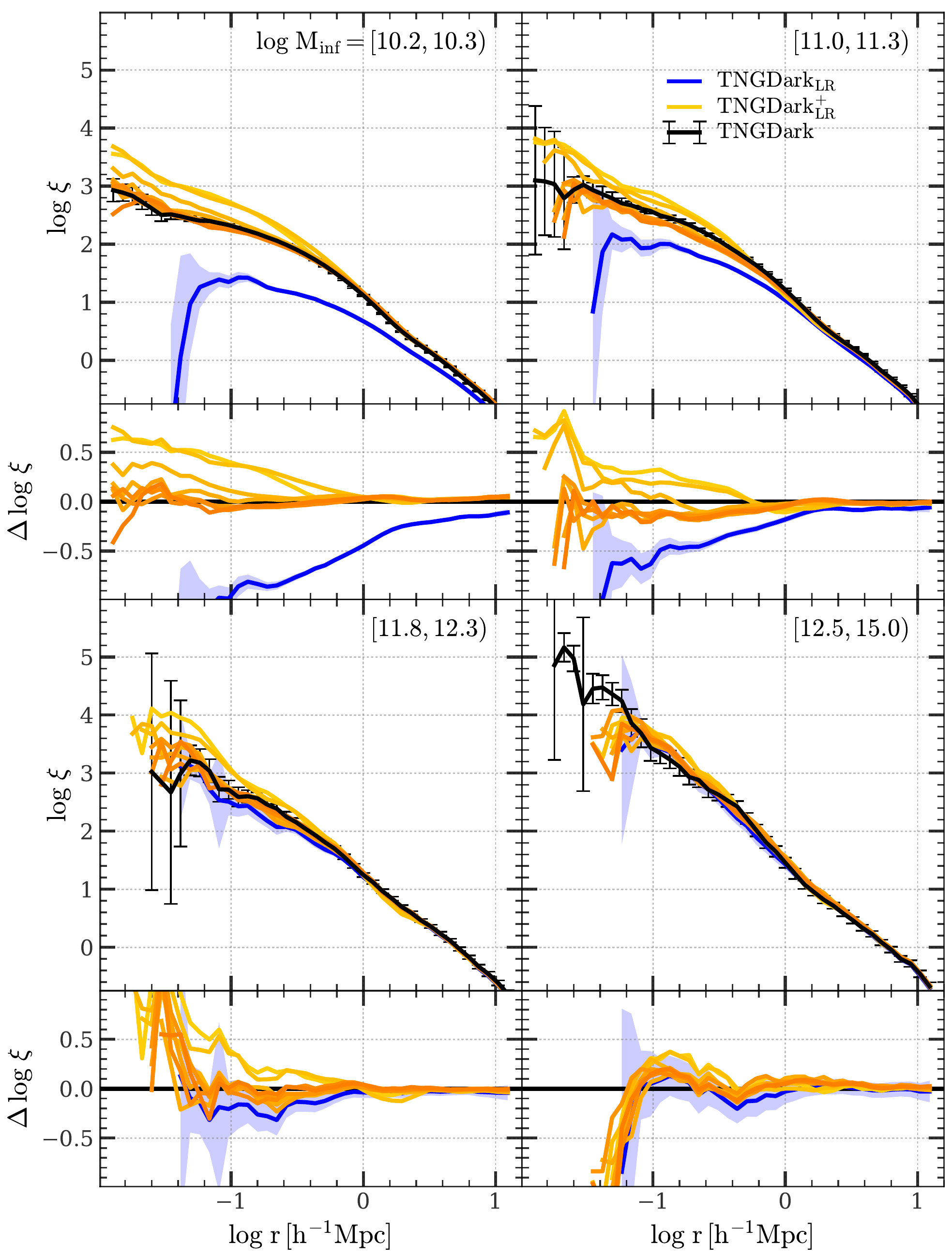}
    \caption{\revstyle
        The same as Fig.~\ref{fig:app-mf-vol-variants}, but here we show the
        two-point correlation functions of all subhalos (central and satellite) 
        at $z=0$ in four different ranges of infall masses. }
    \label{fig:app-tpcf-vol-variants}
\end{figure*}

{\revstyle
As outlined in \S\ref{ssec:outline-algorithm}, the extension algorithm 
operates on branches of a target simulation $S$ at low resolution, requiring that 
each branch includes at least one resolved central subhalo. The completeness 
of the extended trees is thus constrained by this requirement. On the other hand, 
to ensure applicability to all kinds of halos in $S$, the reference simulation 
$S'$ at high resolution must encompass a representative population of halos 
in terms of mass, environments, and assembly histories within the universe. 
The volume size of $S'$ must meet these requirements.

Prior to applying the extension algorithm to a specific target simulation, 
it is imperative to conduct tests that quantify the completeness of the output 
trees from $S$ and verify the fulfillment of requirements for $S'$ in terms of 
desired summary statistics. In this appendix, we provide an example of such tests 
employing a pair of N-body simulations: $S=\,$TNG100-3-Dark 
(referred to as $\rm TNGDark_{LR}$) and $S'={\rm TNGDark}$. 
$\rm TNGDark_{\rm LR}$ serves as a low-resolution counterpart of TNGDark, 
sharing the same box size but possessing a lower mass resolution of 
$m_{\rm dark\ matter}=3.84\times 10^8 \msun$ comparable to that of ELUCID. 
The specific choices of variables and parameters are the same as those employed 
in \S\ref{ssec:specific-elucid-and-tngdark}. Given that these two simulations 
have identical cosmology and initial condition, we can assess the limitations 
and requirements of the extension algorithm itself, unaffected by 
discrepancies in cosmology and volume sampling.
}

\subsection{Completeness of the Extended Population}
\label{app:ssec:completeness}

{\revstyle
Fig.~\ref{fig:app_mfs_completeness_result} shows the infall mass 
functions of satellite subhalos at four different redshifts, obtained from the 
target simulation $S={\rm TNGDark}_{\rm LR}$ using the same method as
in Fig.~\ref{fig:extension-hmf}. By comparing the extended population (gray lines) 
with the simulated one (blue lines), it is evident that the low-mass end of the 
mass function is significantly elevated at each redshift. However, when 
compared to the results obtained from the high-resolution simulation TNGDark 
(black lines), the extended mass functions are still lower by 0.05 (0.15) dex 
at $z=0$ ($z=5$). This incompleteness becomes apparent at infall masses of 
$\sim 10^{11} \msun$ and increases as the mass decreases to the 32-particle 
resolution limit of ${\rm 10^{10.1}} \msun$. This discrepancy arises directly 
from a limitation of the extension algorithm: it is unable to generate 
a branch when the entire central stage is unresolved by $S$. 
The extension algorithm should, therefore, be used with caution 
when these limitations are of critical importance to the application, 
particularly for subhalos with infall masses that approach the resolution limit 
of the target simulation.
Alternatively, deep learning-based super-resolution techniques, such as those 
proposed by \citet{liAIassistedSuperresolutionCosmological2021} and 
\citet{niAIassistedSuperresolutionCosmological2021}, offer a potential solution to 
the problem of incompleteness in unresolved subhalos.
Nonetheless, it is important to note that such methods currently 
only apply to individual snapshots and are incapable of 
recovering assembly histories of unresolved subhalos. Thus, a potential solution
is to perform these methods at a given snapshot of the low-resolution 
simulation, reaching the desired mass limit, statistically match the super-resolved 
subhalos to those with well-resolved histories in a high-resolution simulation, 
and integrate these histories back into the low-resolution simulation. 
This approach needs further exploration.
Above $10^{11.5} \msun$ (equivalent to $\sim 1000$ particles), the extended mass functions 
are in good agreement with those derived from TNGDark at all redshifts.
This indicates that unresolved branches do not affect the completeness of 
the extended population with mass above this threshold.
}

\subsection{Convergence of the Algorithm}
\label{app:ssec:convergence}

{\revstyle
To determine the required volume size of the reference simulation, we apply
the extension algorithm to $\rm S= TNGDark_{\rm LR}$ with a series of subvolumes 
in $\rm S'=TNGDark$ of different sizes. These subboxes have side lengths 
of $L_{\rm sub}=20$, $25$, $32$, $40$, $50$, and $60\,\mpc$, respectively.
The obtained results for each subbox are compared to those of the full box with 
$L_{\rm sub}=L_{\rm box}=75\,\mpc$. The chosen subboxes correspond to 
fractions $f_{\rm sub} = 2\%$, $4\%$, $8\%$, $15\%$, $30\%$, $51\%$ and $81\%$ 
of the full volume. The infall mass functions of satellite subhalos at $z=0$ 
are presented in Fig.~\ref{fig:app-mf-vol-variants}, while the TPCFs of all subhalos, 
both central and satellite, in different mass bins are shown in 
Fig.~\ref{fig:app-tpcf-vol-variants}.

It is seen that the mass function of the extended population remains stable 
regardless of the volume size of the reference simulation. The difference in 
mass functions between the smallest subbox ($f_{\rm sub}=2\%$) and the full box 
is less than 0.1 dex for $M_{\rm inf} < 10^{12} \msun$, with the algorithm 
demonstrating convergence when $f_{\rm sub} \ge 15\%$. However, for higher-mass 
subhalos ($M_{\rm inf} \geqslant 10^{12} \msun$), fluctuations are more evident 
in both the mass functions themselves and the differences between them. 
This is due to the rarity of massive (sub)structures within a limited volume. 
Thus, for such massive subhalos, a larger subvolume of $S'$ yields a more 
unbiased result.

The analysis of the higher-order statistic, TPCF, is more complex. 
When using a subbox with $f_{\rm sub}=2\%$, the TPCF of the extended population 
significantly overestimates the clustering of subhalos of all masses at 
$r < 0.5 \mpc$. This overestimation is more significant for lower-mass subhalos 
and smaller halo-centric distances, where the extension algorithm needs to create 
more subhalos. As the subvolume of $S'$ increases, the TPCF of $S$ becomes more 
similar to that of TNGDark and converges at $f_{\rm sub} \ge 15\%$.

Based on these tests, we can conclude that for a target mass resolution comparable 
to $\rm TNGDark_{LR}$ and the summary statistics considered here, a subvolume of 
$L_{\rm sub} \sim 40\mpc$ (approximately 15\% of the volume of TNGDark) is 
marginally sufficient for the algorithm to function properly. As a result, 
using TNGDark as the reference simulation is a reliable choice for extending 
ELUCID.
}

\bsp	
\label{lastpage}
\end{document}